\documentclass[fleqn,usenatbib]{mnras} 
\usepackage{newtxtext,newtxmath}  
\usepackage{amsmath}
\usepackage{graphicx}
\usepackage{float}
\usepackage{placeins}

\usepackage{color}

\newcommand{\equ}[1]{eq.~(\ref{eq:#1})}

\newcommand{\equnp}[1]{eq.~\ref{eq:#1}}
\newcommand{\se}[1]{\S\ref{sec:#1}}
\newcommand{\fig}[1]{Fig.~\ref{fig:#1}}

\newcommand{\Fig}[1]{Figure~\ref{fig:#1}}

\newcommand{\be}{\begin{equation}}
\newcommand{\ee}{\end{equation}}
\newcommand{\ba}{\begin{align}}
\newcommand{\ea}{\end{align}}
\newcommand{\bad}{\begin{equation} \begin{aligned}}
\newcommand{\ead}{\end{aligned} \end{equation}}
\newcommand{\bea}{\begin{eqnarray}}
\newcommand{\eea}{\end{eqnarray}}
\def\ra{\rangle}
\def\la{\langle}

\newcommand{\bul}{$\bullet\ $}

\newcommand{\no}{\noindent}

\newcommand{\msun}{M_\odot}
\newcommand{\Msun}{M_\odot}

\newcommand{\lsun}{L_\odot}

\newcommand{\Zsun}{Z_\odot}
\newcommand{\Zsolar}{Z_\odot}
\newcommand{\ifm}[1]{\relax\ifmmode#1\else$\mathsurround=0pt #1$\fi}
\newcommand{\kms}{\ifmmode\,{\rm km}\,{\rm s}^{-1}\else km$\,$s$^{-1}$\fi}

\newcommand{\Mpc}{\,{\rm Mpc}}
\newcommand{\kpc}{\,{\rm kpc}}
\newcommand{\pc}{\,{\rm pc}}
\newcommand{\cm}{\,{\rm cm}}
\newcommand{\Gyr}{\,{\rm Gyr}}

\newcommand{\Myr}{\,{\rm Myr}}
\newcommand{\myr}{\,{\rm Myr}}

\newcommand{\yr}{\,{\rm yr}}
\newcommand{\erg}{\,{\rm erg}}
\newcommand{\ergs}{\,{\rm erg}\,{\rm s}^{-1}}

\newcommand{\cmc}{\,{\rm cm}^{-3}}

\newcommand{\ltsima}{$\; \buildrel < \over \sim \;$}
\newcommand{\lsim}{\lower.5ex\hbox{\ltsima}}
\newcommand{\gtsima}{$\; \buildrel > \over \sim \;$}
\newcommand{\gsim}{\lower.5ex\hbox{\gtsima}}
\newcommand{\prop}{\propto}

\def\omm{\Omega_{\rm m}}
\def\oml{\Omega_{\Lambda}}

\def\Mv{M_{\rm v}}

\def\Mveight{M_{{\rm v},10.8}}

\def\Rv{R_{\rm v}}
\def\Vv{V_{\rm v}}
\def\Tv{T_{\rm v}}

\def\nbv{n_{\rm bv}}
\def\nb{n_{\rm b}}
\def\tv{t_{\rm v}}

\def\Mg{M_{\rm g}}
\def\Ms{M_{\rm s}}

\def\Re{R_{\rm e}}
\def\fg{f_{\rm g}}
\def\fgs{f_{\rm gs}}
\def\Rfilv{R_{{\rm fil},{\rm v}}}

\def\Sig1{\Sigma_1}

\def\Mt{M_{\rm tot}}
\def\Mtot{M_{\rm tot}}

\def\Md{M_{\rm d}}
\def\Rd{R_{\rm d}}
\def\Hd{H_{\rm d}}

\def\Mc{M_{\rm c}}
\def\Rc{R_{\rm c}}

\def\Vc{V_{\rm c}}

\def\td{t_{\rm d}}

\def\tmer{t_{\rm mer}}

\def\fb{f_{\rm b}}

\def\brho{\bar\rho}

\def\Rs{R_{\rm str}}

\def\Mbh{M_{\rm bh}}

\def\Mj{M_{\rm J}}

\def\eps2{\epsilon_{-2}}
\def\mp{m_{\rm p}}
\def\tff{t_{\rm ff}}

\def\tfbk{t_{\rm fbk}}
\def\tcool{t_{\rm cool}}
\def\tvir{t_{\rm vir}}
\def\eps{\epsilon}
\def\tcc{t_{\rm cc}}

\def\ssim{\!\sim\!}
\def\seq{\!=\!}
\def\ssimeq{\!\simeq\!}

\def\sgt{\!>\!}
\def\slt{\!<\!}
\def\sgsim{\!\gsim\!}
\def\slsim{\!\lsim\!}
\def\sgeq{\!\geq\!}
\def\sleq{\!\leq\!}
\def\sgg{\!\gg\!}
\def\sll{\!\ll\!}
\def\sdash{\!-\!}
\def\stimes{\!\times\!}
\def\sprop{\!\propto\!}

\def\Rethree{R_{{\rm e},0.3}}
\def\Mach{{\mathcal M}}

\def\Rsh{R_{\rm sh}}
\def\Rshone{R_{{\rm sh},1}}

\def\Rsseven{R_{{\rm str},0.7}}
\def\nsh{n_{\rm sh}}
\def\mp{m_{\rm p}}

\def\Mj{M_{\rm J}}
\def\Rj{R_{\rm J}}
\def\Vw{V_{\rm w}}
\def\Rc{R_{\rm c}}
\def\rhoc{\rho_{\rm c}}
\def\rhow{\rho_{\rm w}}

\def\nin{n_{\rm str}}
\def\rhoin{\rho_{\rm str}}
\def\rhosh{\rho_{\rm sh}}
\def\tw{t_{\rm w}}
\def\Mcsix{M_{{\rm c},6}}

\def\MT{M_{\rm T}}
\def\RT{R_{\rm T}}
\def\Mdotac{\dot{M}_{\rm ac}}
\def\Mdotv{\dot{M}_{\rm v}}
\def\Mdotw{\dot{M}_{\rm w}}

\def\nobs{n_{\rm gal}}
\def\ndisc{n_{\rm d}}
\def\ncool{n_{\rm cool}}
\def\nfbk{n_{\rm fbk}}
\def\nsh{n_{\rm sh}}

\def\sigd{\sigma_{\rm d}}
\def\Vd{V_{\rm d}}
\def\Sigd{\Sigma_{\rm d}}

\def\Rg{R_{\rm gal}}

\def\Rgone{R_{{\rm gal},1}}
\def\Mgen{M_{\rm gen}}
\def\Mgennine{M_{{\rm gen},9}}
\def\Mgensix{M_{{\rm gen},6}}
\def\Mcsix{M_{{\rm c},6}}
\def\cs{c_{\rm s}}
\def\Mcshield{M_{\rm c,shield}}

\def\Vwthree{V_{{\rm w},3.5}}

\title[Feedback-free starbursts at cosmic dawn]
{Efficient Formation of Massive Galaxies at Cosmic Dawn 
by Feedback-Free Starbursts 
}

\author[Dekel et al.]
{\parbox[t]{\textwidth}
{Avishai Dekel$^{1,2}$\thanks{E-mail: dekel@huji.ac.il},
Kartick C. Sarkar$^{1,3}$,
Yuval Birnboim$^1$,
Nir Mandelker$^1$,
Zhaozhou Li$^1$
}
\\ \\ 
$^1$Racah Institute of Physics, The Hebrew University, Jerusalem 91904 Israel\\
$^2$SCIPP, University of California, Santa Cruz, CA 95064, USA\\
$^3$Astronomy, Tel Aviv University, Tel Aviv, Israel\\
}

\begin{document}

\large  

\pagerange{\pageref{firstpage}--\pageref{lastpage}} \pubyear{2002}

\maketitle

\label{firstpage}

\begin{abstract}
JWST observations indicate a surprising excess of luminous galaxies at 
$z\ssim 10$ and above, consistent with efficient conversion of the accreted 
gas into stars, unlike the suppression of star formation by feedback at later 
times.
We show that the high densities and low metallicities at this epoch
{\it guarantee\,} a high star-formation efficiency (SFE) in the most massive 
dark-matter haloes. 
Feedback-free starbursts (FFBs) occur when the free-fall time is 
shorter than $\sim\!1\Myr$, below the time for low-metallicity massive stars 
to develop winds and supernovae.  
This corresponds to a characteristic density of $\sim\! 3\stimes 10^3\cmc$.
A comparable threshold density permits a starburst by allowing cooling to 
star-forming temperatures in a free-fall time.
The galaxies within $\sim\!10^{11}\msun$ haloes at $z \ssim 10$ are 
expected to have FFB densities.
The halo masses allow efficient gas supply by cold streams in a halo 
crossing time $\sim\!80\Myr$.
The FFBs gradually turn all the accreted gas into stars in clusters
of $\sim\! 10^{4\sdash 7}\msun$ within galaxies that are rotating 
discs or shells.     
The starbursting clouds are insensitive to radiative feedback and
are shielded against feedback from earlier stars.
We predict high SFE above thresholds in redshift and 
halo mass, where the density is $10^{3\sdash 4}\cmc$. 
The $z\ssim 10$ haloes of $\sim 10^{10.8}\msun$ are predicted to host 
galaxies of $\sim\! 10^{10}\msun$ with SFR $\sim\!65\msun\yr^{-1}$, 
blue colors, and sub-kpc sizes.     
The metallicity is $\leq\! 0.1\Zsun$ with little dust, gas, outflows and hot 
circum-galactic gas, allowing a top-heavy IMF but not requiring it. 
The compact galaxies with thousands of young FFB clusters
may have implications on reionization, black-hole growth and globular 
clusters.
\end{abstract}


\section{Introduction}
\label{sec:intro}

A puzzling excess of bright galaxies is indicated in the first samples of 
high-$z$ galaxies from {\tt NIRCam} aboard {\tt JWST}
\citep[e.g.,][]{naidu22,haslbauer22,
finkelstein22b,finkelstein23,donnan23a,donnan23b,labbe23,
bouwens23,mason23,lovell23,adams23,perez23,arrabal-haro23}.
It is an order-of-magnitude excess in the abundance of the brightest 
galaxies at (partly photometric) redshifts $z\seq 7\sdash 16$ 
(hereafter generally $z\ssim 10$)
compared to the expectations based on the common wisdom of galaxy 
formation within the standard cosmological paradigm of $\Lambda$CDM
\citep[e.g.,][]{boylan23,finkelstein23,wilkins23,yung23}. 
This motivates studies of the possible distinguishing features
of galaxy formation at these early epochs compared to the more
familiar galaxies at lower redshifts.
Our study aims at first-principle physical processes
rather than an ad-hoc attempt to explain the apparently 
surprising and still uncertain observations.

\smallskip 
A rather exotic explanation would be a deviation from the standard cosmology,
in which the number density of dark-matter haloes above a given 
mass at $z \ssim 10$ is larger than in $\Lambda$CDM. 
An example of a model that could possibly provide such a deviation 
is the addition of ``Early Dark Energy" (EDE) to $\Lambda$CDM \citep{klypin21}, 
originally proposed as a solution to the other puzzle of ``Hubble tension" 
\citep{karwal16,poulin19,kamionkowski22}. 
Another example of an ad-hoc idea is of an accelerated growth of early
structure via extremely massive primordial black-hole seeds \citep{liu22}.

\smallskip 
Within standard $\Lambda$CDM, one could in principle appeal to 
an ad-hoc top-heavy 
stellar initial mass function (IMF), consisting of a sufficiently large 
fraction of massive stars with high UV luminosities 
\citep{zackrisson11,inayoshi22,steinhardt22,harikane23}. 
This would increase the luminosities without requiring excessively large 
stellar masses.  Qualitatively, a top-heavy IMF is possible if the 
metallicity is sufficiently low 
and if the gas temperature is sufficiently high. 
In this case,
the challenges would be to have the top-heavy IMF explain (a) a full
order-of-magnitude increase in the luminosity-to-mass ratio 
\citep{yajima22,papovich23,perez23}, 
(b) metallicities that are not sufficiently low \citep{arellano22,
langeroodi22,schaerer22,trump23,curtis23,fujimoto23,katz23},
(c) low dust attenuation that may be in tension with excessive supernova 
feedback \citep{fiore23},
and (d) a mass dependence of the luminosity excess, 
if such a dependence is detected.
Excessive UV due to radiation from central black holes
may be relevant only if active galactic nuclei (AGN) turn out to be
abundant in these galaxies (\se{obs}).

\smallskip 
We show below that at $z\ssim 10$, and especially at large halo masses,
the efficiency of conversion of the gas accreted onto the haloes to stars in 
the central galaxies is {\it expected\,} to be high, as a direct result of the 
high cosmological density at these epochs and the expected low metallicity.
For a given halo mass $\Mv$ and stellar mass $\Ms$, the SFE $\eps$
is defined by 
\be 
\Ms = \eps\, \fb\, \Mv \, ,
\ee
where $\fb \ssimeq 0.16$ is the universal baryonic fraction.
An analogous efficiency can be defined by relating the
star-formation rate (SFR) to the gas accretion rate onto the halo,
\be
{\rm SFR} = \eps'\, \Mdotac \, .
\ee
We show that the massive haloes at $z \ssim 10$
are naturally expected to have $\eps \ssim 1$, as opposed to galaxies at 
later times or lower masses which show $\eps \ssim 0.1$ or lower
values based on abundance matching of stellar masses and $\Lambda$CDM haloes
\citep{rodriguez17,moster18,behroozi19}.
This low efficiency is assumed to be the result of {\it feedback\,} from stars 
and supernovae (or AGN at large masses), 
which suppresses the star formation rate by heating the gas or 
ejecting it from the star-forming regions. 
The key for a high $\eps$ is therefore that the star formation be free
of the effects of feedback.
In addition, the gas supply should not be inhibited by a hot circum-galactic 
medium (CGM), and all of the gas should participate in the fragmentation
to star-forming clouds.

\smallskip 
The requirement of gas supply through the halo is fulfilled because even haloes 
of $\sim\! 10^{11}\msun$ were in the cold-flow regime, 
where the atomic cooling time down to temperatures of $\sim\!10^4$K
is much shorter than the halo crossing time
(\se{cold_inflow}).
These haloes are below the threshold mass of 
$\sim\! 10^{11.5}\msun$ for virial shock heating, only above 
which the CGM is expected to contain an extended hot component 
\citep{db06,fielding17,stern21}.

\smallskip 
The key feature addressed in the current paper is that feedback-free star 
formation is expected when the density is above a critical value  
$\sim\! 3 \stimes 10^3\cmc$. 
At these densities the free-fall time is $<\! 1\Myr$, 
shorter than the delay time between a starburst and the onset of 
effective feedback by supernovae, as well as
by stellar winds once the metallicity is below $\sim\! 0.2\Zsolar$.  
If the starbursts occur in clouds of $\geq\!10\pc$, the surface density
is above the threshold that prevents radiative feedback from significantly
perturbing the FFB process.

\smallskip 
In order to enable a coherent FFB period, the star formation must 
occur in a free-fall burst rather than gradually during 
the inflow over a halo crossing time, $\tvir \ssim 80\Myr$ at $z\ssim 10$.
For this purpose we address a threshold for star formation due to the need to 
cool below $10^4$K on a free-fall time, and show that it is fulfilled at a 
density which is only slightly above the FFB density.
Furthermore, for FFB we should verify that the starbursting clouds, once
above a threshold mass, are self-shielded against winds and radiation from 
neighboring star clusters of earlier generations.

\smallskip 
A second key point is that the densities for FFB are expected
to arise naturally at $z\ssim 10$, especially in massive haloes.
We address below the density according to two extreme scenarios. 
One, in which the supersonic streams that feed the galaxies are largely radial 
and shock at the galaxy boundary.  The other, where the streams of 
somewhat higher angular momentum settle into discs whose sizes are determined 
by angular momentum.

\smallskip 
We then investigate the starbursting clouds in the two scenarios. 
In the radial-stream scenario the clouds are associated with the Jeans 
mass in a thin shell, which is supported from the interior 
by a cumulative hot supernova bubble and from the exterior by
the inflowing streams.
In the disc scenario the clouds are associated with the Toomre mass 
under marginal instability. In both cases, the clumps form
in generations that are determined by the need to accumulate sufficient
mass by accretion; 
either a Jeans mass in a Jeans area on the shell,
or sufficient disc mass for making the disc unstable with a Toomre parameter 
$Q \ssim 1$.
We attempt below to estimate the expected radius of the cloud-forming shell or
disc, which will tell how compact the galaxies are expected to be.

\begin{figure*} 
\centering
\includegraphics[width=0.90\textwidth,trim={0.0cm 0.2cm 0.0cm 0.2cm},clip]
{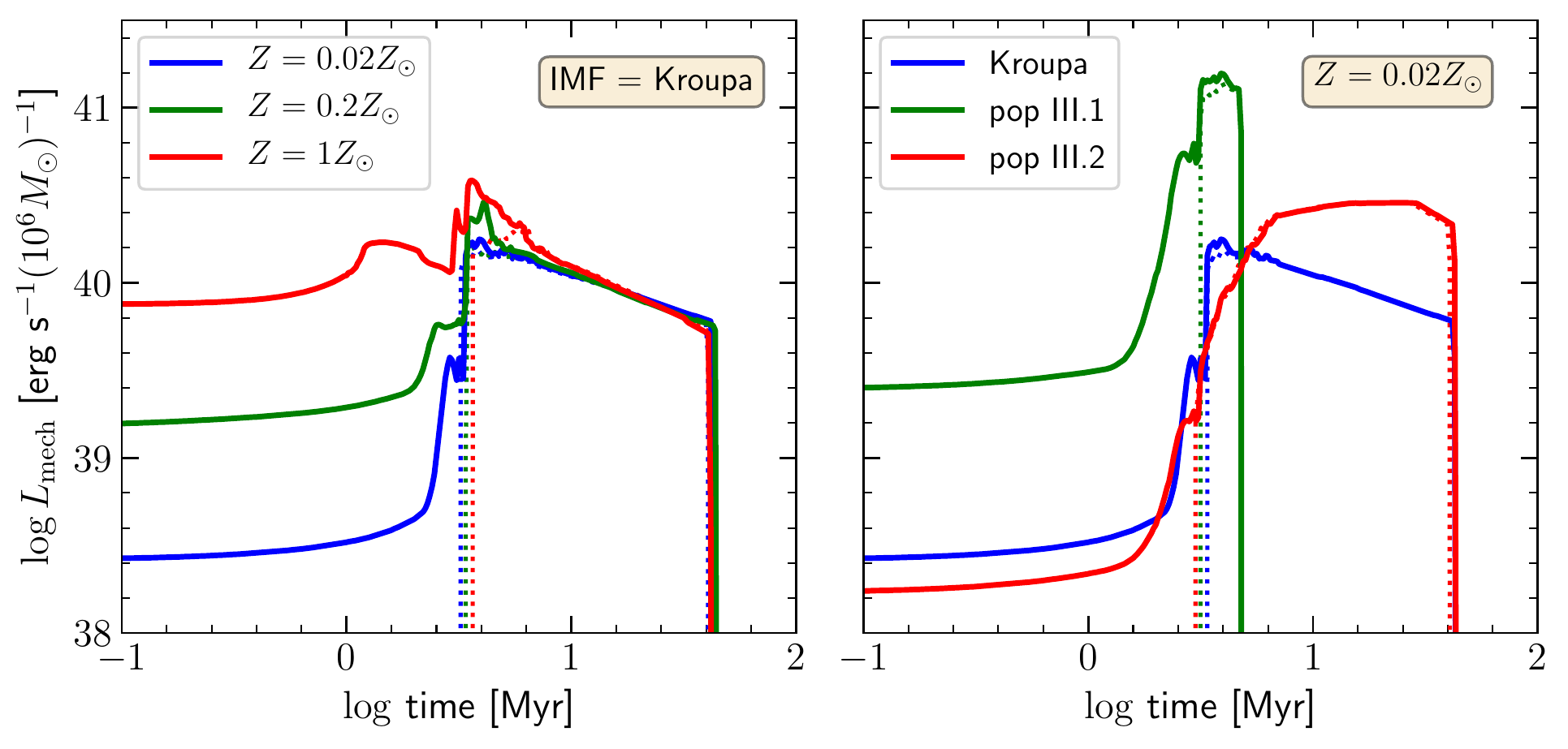} 
\caption{
Feedback mechanical energy injection rate as computed by {\tt Starburst99}
for an instantaneous starburst in a star cluster of $10^6\msun$
as a function of time from the burst.
Shown are curves for different metallicities and different IMFs.
For each case, the solid curve refers to the total energy of stellar wind and
supernova and the dotted curve is the contribution of supernovae only.
{\bf Left:} Three different metallicities for a standard Kroupa IMF.
{\bf Right:} Three different IMFs at a low metallicity $Z \seq 0.02\Zsun$.
Robustly in all cases, the onset of supernova feedback is sharp at
$t\ssimeq 3\Myr$ when it becomes dominant.
Except for $Z \seq 1\Zsun$, the early stellar-wind feedback is negligible
until it rises steeply near $t \ssimeq 2\Myr$.
Thus, at sufficiently low metallicity,
a burst of $\sim\! 1\Myr$ is expected to be safely free of both stellar-wind
feedback and supernova feedback.
}
\vspace{-10pt}
\label{fig:sb99_Lmech}
\end{figure*}

\smallskip 
Based on the halo mass function according to the standard $\Lambda$CDM 
cosmology, 
the dark-matter halo mass for each observed galaxy can be estimated rather 
robustly for a complete sample in a given effective volume.
The effective stellar radii can also be observed quite robustly.
On the other hand, the stellar masses, as derived from SED fitting,
are more uncertain as they are based on the mass-to-light ratio 
that is associated with an assumed IMF and star-formation history (SFH). 
The preliminary results from {\tt NIRCam} will be complemented by upcoming
data from {\tt MIRI} and {\tt NIRSpec}.
We sort out below observable predictions to be compared to current and future
{\tt JWST} observations. These should first be the redshift and mass 
dependence of $\eps$, via the associated stellar masses and SFR,
and the characteristic density of $3\stimes 10^3\cmc$.
Other observables to be addressed include the compact sizes, low metallicities, 
low gas fraction and low dust attenuation, the cold streams with low levels 
of outflows and hot CGM, and the morphology of clumpy discs or clusters in 
shells.  
We then discuss possible implications of the FFB scenario.
One is an efficient post-FFB growth of seed black holes due to the intense
interactions of massive stars within the young clusters followed by
cluster-cluster mergers in the compact galaxies.
Another is the abundance of globular clusters at later times.

\smallskip
This paper is organized as follows.
In \se{fbk} we quantify the feedback-free period following a low-$Z$ starburst.
In \se{burst} we propose a possible origin for such a starburst at the critical
density for proper cooling.
In \se{shield} we address the shielding of clouds above a given mass
against winds and UV from earlier generations of star formation.
In \se{cold_inflow} we discuss the cold inflow through the halo.
In \se{density} we work out the expected characteristic densities at 
$z \ssim 10$ in the two scenarios, of discs and of radial inflow.
In \se{clusters} we study he fragmentation into star-forming clouds in
several generations.
In \se{shells} we estimate the shell radius in the radial inflow scenario.
In \se{obs} we describe current and future comparisons with {\tt JWST}
observations.
In \se{discussion} we discuss certain elements of our results and possible
post-FFB implication.
Finally, in \se{conc} we summarize our conclusions.

\section{Feedback-free Time and Density}
\label{sec:fbk}

\smallskip 
Energy considerations indicate that supernova feedback is capable of 
suppressing SFR in galaxies where the virial velocity is $\Vv \ssim 100\kms$ 
or lower \citep{ds86}.
From \equ{Vv} below, 
at $z \ssim 10$ this corresponds to a virial
halo mass of a few $10^{10}\msun$. 
Since the halo masses in question are not larger than this critical mass by
much, the SFR in them and in their star-forming subsystems 
\citep{dekel23_gclumps}
could in principle be suppressed by feedback, as it is suppressed at lower 
redshifts.
Having no hot CGM in these haloes is expected to allow the supernova bubbles 
escape from the galaxy and thus be more effective in suppressing star 
formation \citep{fielding17}. 
However, as demonstrated below, a burst of star formation at low
metallicities is followed by a delay in the onset of effective
feedback on the order of
\be
\tfbk \sgsim 1\Myr\, .
\label{eq:tfbk}
\ee
If the free-fall time in the star-forming region is shorter than $\tfbk$,
the star formation can proceed freely without suppression even near the
critical halo mass for effective feedback.
The free-fall time is related to the gas number density by
\be
\tff = \left( \frac{3\pi}{32 G \rho} \right) ^{1/2}
= 0.84 \Myr\, n_{3.5}^{-1/2} \, ,  
\label{eq:tff}
\ee
where $n \seq 10^{3.5}\cmc\,n_{3.5}$, measured in units of $\mu \mp$, where
$\mp$ is the proton mass. 
We adopt hereafter a mean molecular weight of $\mu \seq 1.2$, appropriate for 
neutral Hydrogen and Helium gas at $T \sleq 10^4$K.
The threshold of $1\Myr$ then implies a lower bound on the gas density that 
would allow feedback-free star formation,
\be
n > \nfbk = 2.23 \times 10^3 \cmc \, .  
\label{eq:nfbk}
\ee

\subsection{Supernovae and stellar winds}
\label{sec:sn_wind}

\smallskip 
The feedback delay is straightforward for supernova feedback, given the finite
lifetime of the massive stars that eventually dominate the supernova energy.
It turns out that
when the metallicity is low, a similar delay appears also for the
preceding stellar winds.
These delays are demonstrated in \fig{sb99_Lmech},
which shows the evolution of the mechanical
energy rate from an instantaneous starburst as computed using {\tt Starburst99}
\citep{leitherer99} with different metallicities and IMFs.

\smallskip 
The metallicity in \fig{sb99_Lmech} ranges from solar to 0.02 solar 
and it can be qualitatively extrapolated to lower values.
The fiducial IMF assumed is the Kroupa IMF \citep{kroupa01},
a power law with a slope $\alpha\seq -2.35$ in the mass range 
$[0.1-100]\msun$.
The two top-heavy IMFs in the right panel
are extreme examples used for primordial pop III stars
\citep[][the Yggdrasil model]{zackrisson11}.
The Pop-III.1 stars are assumed to form first with a characteristic mass
$\sim\!100\msun$, represented by a power law of $\alpha\seq 2.35$ in
the mass range $[50\sdash 500]\msun$ \citep{schaerer02}.
The Pop-III.2 stars are assumed to form somewhat later with a lognormal
distribution about a characteristic mass $10\msun$ and dispersion
$\sigma \seq 1\msun$, but with wings that extend within $[1-500]\msun$
\citep[][the TA model]{raiter10}.
The lower masses of these stars are assumed to be due to Hydrogen-Deuterium
molecular cooling promoted 
by the Lyman-Werner feedback that was provided by the Pop-III.1 stars
\citep{mackey03}. 

\begin{figure*} 
\centering
\includegraphics[width=0.90\textwidth,trim={0.0cm 0.2cm 0.0cm 0.2cm},clip]
{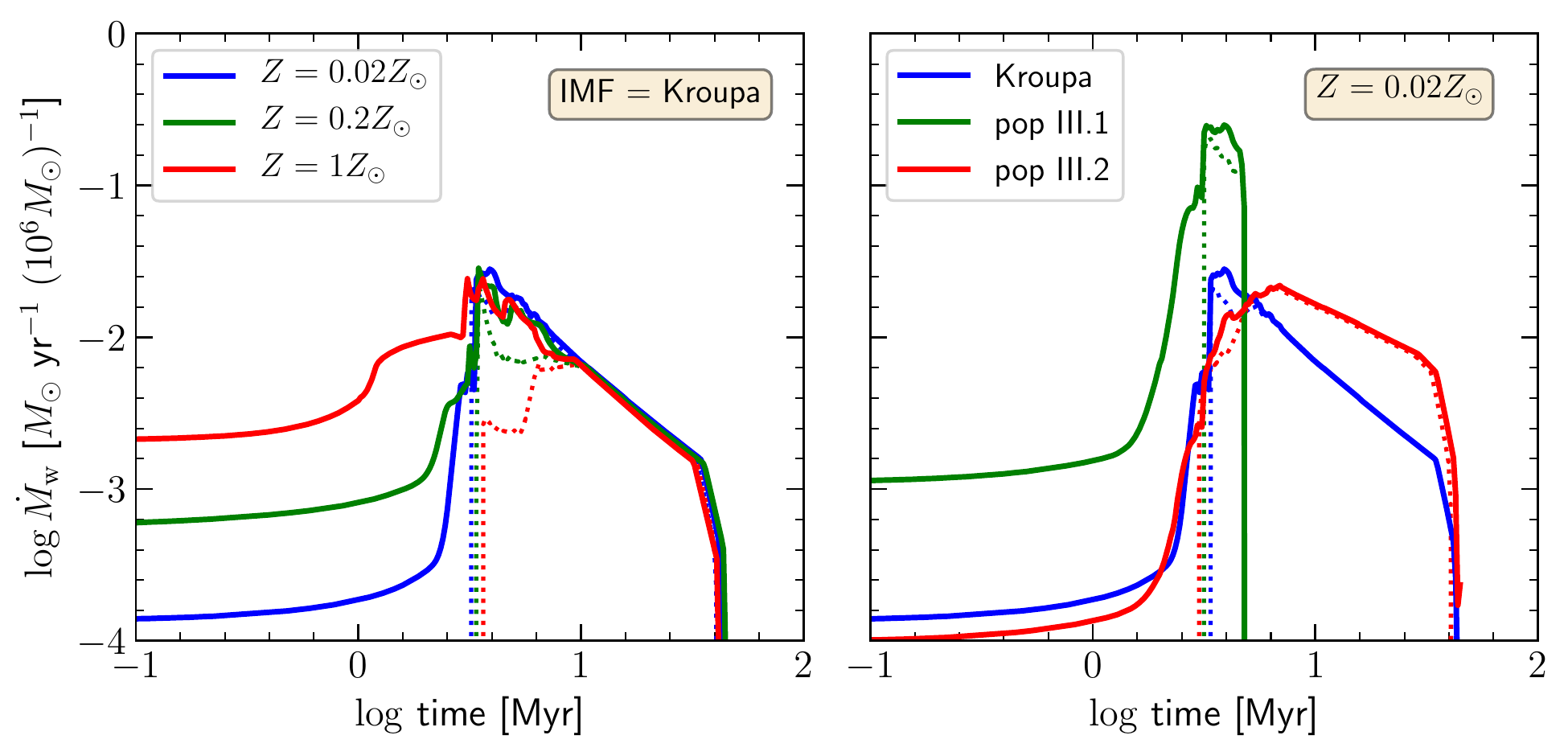} 
\caption{
Mass loss rate as computed by {\tt Starburst99}
for an instantaneous starburst in a star cluster of $10^6\msun$
as a function of time from the burst.
Shown are curves for different metallicities and different IMFs as in
\fig{sb99_Lmech}.
For Kroupa IMF we deduce an effective peak wind of
$\dot{M}_{\rm w} \ssimeq 0.01\msun\yr^{-1}$ per $10^6\msun$,
that lasts for $\tw \ssimeq 10\Myr$.
}
\vspace{-10pt}
\label{fig:sb99_Mdot}
\end{figure*}

\smallskip
\Fig{sb99_Lmech} shows that the delay in supernova feedback is robust near 
$3\Myr$ with only little sensitivity to metallicity or IMF.
This is because the nuclear burning time on the main sequence for massive stars 
is roughly independent of mass due to the Eddington limit,
which imposes a linear relation between luminosity and stellar mass,
$L \sprop M$ (as opposed to $\prop\!M^{3.5}$ for lower mass stars),
and therefore a constant lifetime $\prop\! 0.15 M/L \ssim 3\Myr$.
The stellar winds, on the other hand, are very sensitive to metallicity,
through its effects on the opacity due to many resonant transitions;
for less than solar metallicities the winds have an effective delay of
$\sim\! 2\Myr$.
In a little more detail,
during the first $\sim\!1\Myr$, dominated by stars on the main sequence,
the wind power is $\prop Z^{0.9}$ \citep{hirschi07}, 
so it is rather low for low metallicities.
During $t\seq 1\sdash 2\Myr$, after the original Hydrogen envelope has been
lost by wind, the stellar surface is enriched with metals via convection,
which causes a first significant rise in the wind power.
Core Helium burning starts near $t \ssim 2\Myr$, producing metals such as C, O
and N, which make the wind power boost up to a peak corresponding to the
Wolf-Rayet (WR) phase of the most massive stars. The wind power typically
diminishes after a few Myr, after the less massive stars have ended this phase
\citep[][\S 3.4 and 4, Fig.~2]{hirschi07}.
At the peak, the wind energy could be comparable to the supernova
energy (at high metallicity) or much smaller (at low metallicity).
Since the same massive stars determine both the power of winds and supernova,
the relative contributions of the two is rather insensitive to the IMF.
Since WR stars evolve from progenitors more massive than a certain 
$Z$-dependent threshold, e.g., $\ssim 60\msun$ for $Z \slt 0.1\Zsun$,
the peak level of the wind also depends on the upper cutoff of the mass range
considered for the IMF.
A minor caveat to mention here is that winds from rapidly rotating
metal-poor stars could be significantly stronger \citep{liu21}, but 
their contribution at intermediate metallicities is under debate and their
abundance is unknown.

\smallskip  
In order to estimate how low the metallicity should be for the wind to be 
ineffective in suppressing a starburst, 
we compare the mechanical energy injected by the wind 
from a starbursting cluster to the binding energy of the cluster.
We pick a typical cluster of $\Mc \seq 10^6\msun$ 
based on \se{clusters} below, e.g., the Jeans mass of \equ{Mj}.
The wind energy of power $L\seq 10^{40}\ergs\, L_{40}$ during a period of 
$t \seq 1\Myr\,t_1$ is 
\be
E_{\rm w} = L\,t \simeq 3.1\times 10^{53}\erg \,L_{40}\,t_1\, \eta \, ,
\ee
where $\eta$ is the fraction of the emitted energy that is actually deposited 
within the cluster.
For the binding energy we adopt a cluster radius of 
$\Rc \ssimeq 14 \pc\, c^{-1/3}$, based on the Jeans radius of \equ{Rj} below
and allowing a collapse factor of $c^{1/3}$.
This gives
\be
E_{\rm b} = \frac{1}{2}\frac{G\Mc^2}{\Rc} 
\simeq 3.6\times 10^{51}\erg\, c^{1/3} \, .
\ee
The ratio of energies is thus
\be
\frac{E_{\rm w}}{E_{\rm b}} \simeq 86\,\eta\,c^{-1/3}\,L_{40}\,t_1\, .
\ee
If we adopt $\eta \ssim 0.1$ at the density of $n \ssim 10^3\cmc$ 
\citep[][Fig.~14]{gupta16}, and $c \sgsim 1$ for the early cloud collapse,
we obtain that the wind should be safely ineffective for $L_{40} \sleq 0.1$,
namely once $Z \sleq 0.1\Zsun$ based on \fig{sb99_Lmech} (left).

\smallskip 
We conclude that for such metallicities the total feedback from winds and 
supernovae is expected not to have a significant affect on the SFR 
during the first $\sim\!2\Myr$ after an instantaneous burst.
We conservatively adopt $\tfbk \ssim 1\Myr$ as our fiducial value in order to
allow for a finite duration on the order of a free-fall time for the 
starburst itself.
This leaves a delay of $\ssim 1\Myr$ between the main body of the starburst 
and the onset of efficient feedback.    
It also allows for a possible contribution from proto-stellar jets in the first
free-fall time \citep{appel22,appel23}.

\begin{figure} 
\centering
\includegraphics[width=0.49\textwidth,trim={1.5cm 5.5cm 1.5cm 4.6cm},clip]
{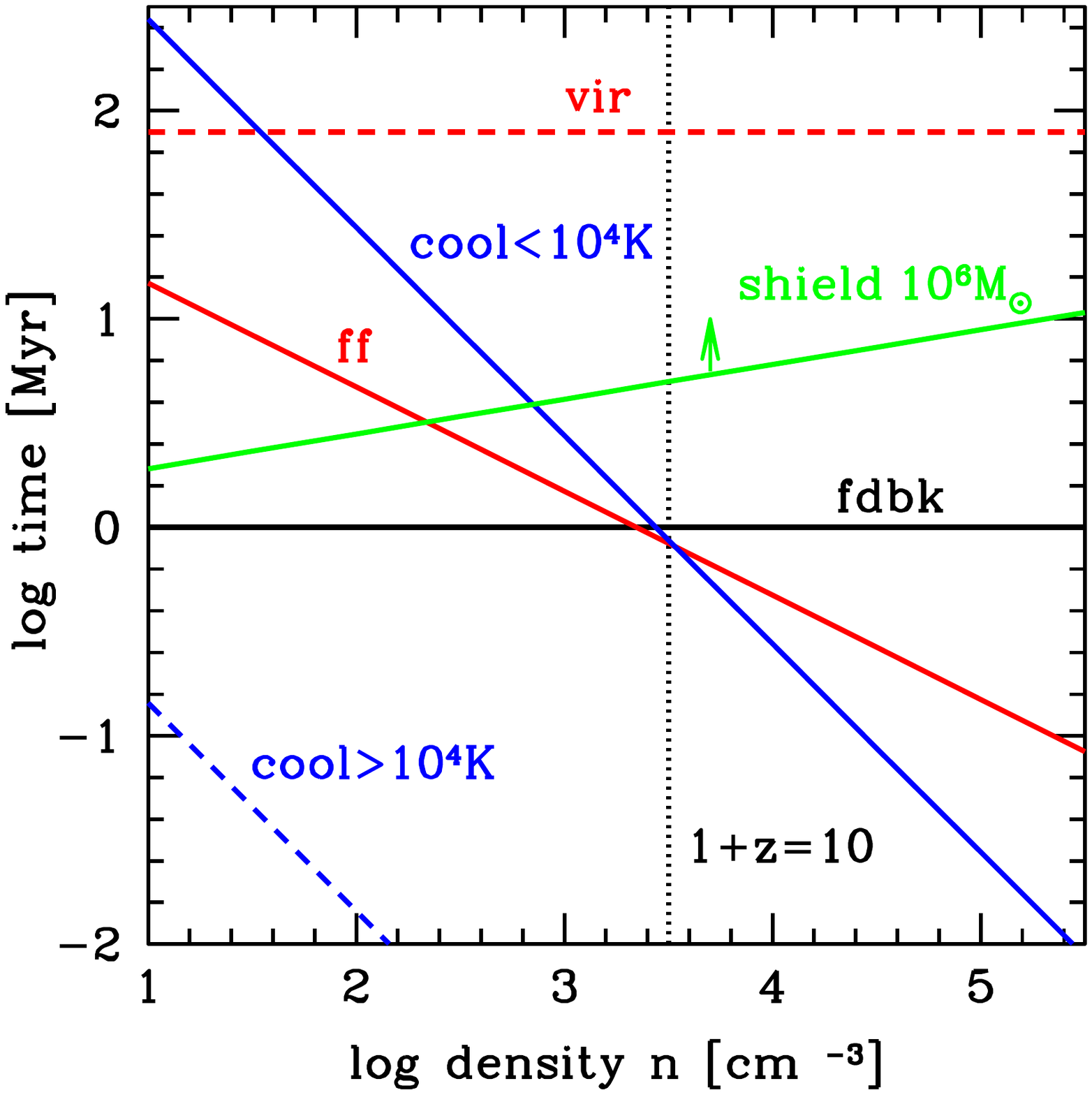} 
\vspace{-15pt}
\caption{
The timescales as a function of density.
The feedback delay time $\tfbk \ssim 1\Myr$ is from \equ{tfbk}.
The free-fall time $\tff$ is from \equ{tff}.
The cooling time at $T\slt 10^4$K $\tcool$ is from \equ{tcool}
with $Z\seq 0.02$ ($\prop\! Z^{-1}$) and $T\seq 10^4$K.
The shielding time is the lower limit for the crushing time $\tcc$
from \equ{tcc2}, for a clump of $\Mc\seq 10^6\msun$ ($\tcc\sprop\! \Mc^{1/3}$)
and winds from a generation of $\Mgen\seq 10^9\msun$
($\tcc\sprop\! \Mgen^{-1/2}$)
in a volume of radius $\Rg\seq 1\kpc$ ($\tcc\sprop\! \Rg$).
The virial time $\tv$ is from \equ{tv} at $1+z\seq 10$
($\prop\! (1+z)^{-3/2}$).
We see that $\tff$ and $\tcool$ are both comparable to $\tfbk\ssim 1\Myr$
or shorter when the density is a few $\times 10^3\cmc$ or larger.
This defines the characteristic density and timescale for feedback-free
starbursts.
The atomic cooling time at $T\sgt 10^4$K
 being much shorter than $\tvir$ and $\tff$ causes
efficient cold inflow through the halo at $T\ssim 10^4$K.
The fact that $\tvir \sgg 1\Myr$ implies gradual formation of
starbursting clumps of the Jeans mass or the Toomre mass
$\sim\!10^6\msun$.
Starbursting clouds above $10^4\msun$ have a shielding time larger than
$1\myr$, indicating that they are shielded against feedback from earlier stars.
}
\vspace{-10pt}
\label{fig:times}
\end{figure}

\smallskip
\Fig{times}, which summarizes the interplay between the relevant timescales and
densities (to be referred to throughout the analysis below), 
illustrates first how the free-fall time drops below the feedback 
time of $\sim\!1\Myr$ at a density $n \sgeq 3\stimes 10^3\cmc$.  

\subsection{Radiative Feedback and Critical Surface Density}
\label{sec:rad}

Beyond the feedback from supernovae and stellar winds,
one should consider whether other sources of feedback
could potentially disturb the feedback-free period.
Of special concern could be radiative stellar feedback, by radiative 
pressure or by photo-ionization.
A general toy model argues that 
feedback that provides a characteristic specific momentum injection rate
$\la \dot{p}/m_\star \ra$ is expected to be incapable of 
overcoming the gravitational binding force and ejecting a significant fraction
of the gas from a star-forming cloud 
once the total surface density $\Sigma_{\rm tot}$ is above a threshold of 
$\Sigma_{\rm crit} \ssim (\pi G)^{-1} \la \dot{p}/m_\star \ra$
\citep{fall10,grudic18,grudic20}.
This is obtained by balancing the forces of feedback,
$\la \dot{p}/m_\star \ra \Ms$, and gravity,
$G M_{\rm tot} M_{\rm gas}/R^2$.
For radiative pressure from a young stellar population one expects
$\la \dot{p}/m_\star \ra \ssim c^{-1} (L/M)_\star 
\ssim 6\times 10^{-8} \cm\, {\rm s}^{-2}$,
assuming $(L/M)_\star \sim 10^3 \lsun/\msun$.
This estimate is confirmed by simulations of star-forming clouds
\citep{grudic18,lancaster21,menon23}.
A similar effective value of $\la \dot{p}/m_\star \ra$ is predicted 
via simulations for photoionization feedback 
\citep{colin13,geen17,kim18}. 
Adopting $\la \dot{p}/m_\star \ra \slsim 10^{-7} \cm\, {\rm s}^{-2}$,
the expected threshold surface density is comparable or slightly smaller than
\be
\Sigma_{\rm crit} \sim 3\stimes 10^3\Msun \pc^{-2} \, ,
\label{eq:Sigma_crit}
\ee
consistent with what is seen in the simulations 
\citep{grudic18,menon23}. 

\smallskip
In particular, by properly solving radiative transfer in high column density 
clouds and combining the effects of direct UV and dust-processed radiation 
pressure, 
\citet{menon23} 
find that near and above this surface density
threshold the SFE is high, 
$\epsilon \ssim 0.7 \sdash 0.8$ by $3\tff$, 
with the rest of the gas partly super-Eddington ejected from the cloud  
toward the end of star formation (as long as $\Sigma \slt 10^5 \msun\pc^{-2}$),
and partly left rotation supported in a thin disc within the cluster. 
This implies that at the end of the FFB phase the clusters are practically
free of interstellar gas, with near-zero column density over most solid angle, 
such that the post-FFB supernova ejecta will escape freely from the clusters,
removing the associated gas and dust (to be used in \se{shield}, \se{shells}
and \se{obs} below).

\smallskip 
The relation between the three-dimensional density and the surface density in a
cloud of diameter $2R$ is 
\be
\frac{\Sigma}{10^{3.5}\msun\pc^{-2}} \ssimeq \frac{n}{10^{3.5}\cmc}\,
\frac{2R}{15\pc} \, .
\label{eq:Sigma_n}
\ee
Thus, in FFB clouds of $n \ssim 10^{3.5}\cmc$, we also expect 
$\Sigma \ssim 10^{3.5}\msun\pc^{-2}$ if the clouds are of radius 
$R \ssim 15 \pc$.
An interesting coincidence is that this size is compatible with the Jeans 
radius that is found in \equ{Rj} below to characterize the initial 
star-forming clouds, 
$\Rj \ssim 14\pc\, n_{3.5}^{-1/2}$. 
For such Jeans clouds, the two densities follow each other,
$\Sigma_{3.5} \ssim n_{3.5}^{1/2}$.
It thus turns out that a Jeans cloud that is free of feedback from stellar
winds and supernovae is also expected not to suffer significant gas ejection 
and the associated suppression of star formation by radiative feedback.

\smallskip 
Once sub-clouds form by non-linear fragmentation,
they would also be largely free of radiative feedback
if they maintain roughly the same surface density as the parent Jeans cloud.
Such a mass-insensitive surface density is valid in local molecular clouds
\citep{krum19} and is seen in cosmological simulations of clumpy discs 
\citep[][Fig.~6]{mandelker17}.
A constant surface density can be qualitatively hinted at  
by combining the relation between velocity dispersion $\sigma$ and radius $R$ 
that characterizes supersonic turbulence, $\sigma \sprop R^{1/2}$, 
with the relation for virial equilibrium within 
clouds of mass $M$, $\sigma^2 \sprop M/R$.

\smallskip 
An additional argument concerning photo-ionization or photo-heating feedback 
is that it is associated with an equilibrium gas temperature $\sim 10^4$K 
\citep{kim18,krum19}, corresponding to a sound speed $\sim\! 10\sdash 15\kms$. 
This indicates that it should be ineffective in driving free winds in clouds 
with escape velocity larger than $15\kms$ \citep{krum19}, at least for gas at 
rest inside the cloud and with no external gas.  Thus, the typical FFB clouds, 
of mass $\sim\!10^6\msun$ and radius $\sim\!10\pc$ (\se{clusters}) such that 
$\Vc\ssim 20\kms$, are expected to be unaffected by photo-ionization feedback. 

\smallskip
Other feedbacks are discussed further in \se{other_feedbacks}.

\bigskip
\section{Cooling density for starbursts}
\label{sec:burst}

A necessary condition for a feedback-free period is that the star formation
in the star-forming clouds occurs in bursts of a short duration, comparable to 
the free-fall time and the feedback-delay time, $\tff \ssim \tfbk \ssim 1\Myr$.
This is a non-trivial challenge, given that the gas supply to the galaxy
by accretion through the halo is much more gradual,
over a timescale on the order of the halo crossing time of 
$\tv \ssim 80 \Myr$, \equ{tv} in \se{cold_inflow} below.
A short starburst following a gas accumulation period can occur if the star 
formation is limited by a threshold in the conditions for star formation, 
in particular in gas density.
One such density threshold, only slightly above $\nfbk$, 
arises from the need for the rather slow cooling time from 
$T \sgsim 10^4$K to the star-formation temperatures of $T \ssim 10$K 
to become shorter than the free-fall time, $\tcool \sleq \tff$
\citep{fernandez18,mandelker18}.
Since for a given temperature and metallicity
$\tcool \sprop \rho^{-1}$ while $\tff \sprop \rho^{-1/2}$, there is a 
critical density above which the relation reverses to $\tcool \sleq \tff$, 
such that collapsing gas clouds can form stars.
We crudely approximate in the $\tcool \sgt \tff$ regime that the temperature
is roughly constant at $T\ssim 10^4$K due to the rapid atomic cooling.

\smallskip 
At metallicities of $Z \sgt 10^{-3} \Zsun$, and at the relevant densities and
temperatures, one dominant cooling process is emission in the
[CII] $158 \mu{\rm m}$ line \citep{krum12_lowZ,pallottini17}.
For a crude estimate, we refer to this as a representative cooling channel,
assuming that the other relevant cooling rates, such as via [OI] and LyA,
are comparable.
The corresponding cooling time is \citep{krum12_lowZ}
\be
\tcool \simeq 0.87\Myr\, n_{3.5}^{-1}\, Z_{0.02}^{-1}\, T_4\, e^{0.009/T_4}\,
C^{-1} \, ,  
\label{eq:tcool}
\ee
where the metallicity (in solar metallicity) is
$Z \seq 0.02\,Z_{0.02}$,
the temperature is $T \seq 10^4{\rm K}\,T_4$,
$\mu$ has been adjusted (from $\mu \seq 1.4$) to $\mu\seq 1.2$,
and $C \seq \la n^2\ra/\la n \ra^2$ is the clumping factor within the
star-forming cloud.
The condition $\tcool \seq \tff$ from \equ{tff} implies a threshold density
\be
\ncool \simeq 3.4\times 10^3\cmc\, Z_{0.02}^{-2}\, T_4^2\, C^{-2} \, .
\label{eq:ncool}
\ee
We note that for $Z$ somewhat higher than $0.02$, this estimate of $\ncool$
becomes lower than $\nfbk$, such that the cooling cannot serve as a threshold
for starburst above $\nfbk$.
However, any partial heating may increase $\tcool$ and therefore $\ncool$ 
to above the estimate in \equ{ncool}, which can make $\ncool \sgt \nfbk$ even 
for the somewhat higher metallicity that may be expected at the advanced
stages of the FFB process (\se{clusters}).
Such partial heating may be provided, e.g.,
by feedback from the earlier generations of FFBs, 
or by turbulence \citep{fernandez18} that is driven by the incoming streams 
\citep{ginzburg22}. 

\smallskip
The crossing of $\tcool$ and $\tff$ at $\ncool$, near and slightly above
the crossing of of $\tff$ and $\tfbk$, is illustrated in \fig{times}.
Starting at $T\ssim 10^4$K with a uniform cloud of $C\ssim 1$, 
the overall density in the 
$z\ssim 10$ massive galaxies (see \se{density} below) may be in the ball 
park of $\ncool$, allowing certain gas accumulation.  
However, once the temperature starts dropping and clumps start 
collapsing, possibly also increasing $C$, 
the density in the clouds becomes significantly
larger than the threshold of $\ncool$ and can thus give rise to a FFB. 
The onset of SFR at $\ncool$ can be rather sharp given that the density in a
clump can be boosted by an order of magnitude during a free-fall collapse.

\section{Shielded clouds}
\label{sec:shield}

\subsection{Shielding against stellar feedback}
\label{sec:shield_fbk}

We assume that the FFBs of $\sim\! 1\Myr$ occur inside clouds of typical mass 
$\Mc \ssim 10^6\msun\,\Mcsix$, to be estimated in \se{clusters} below.
They occur gradually over a longer period comparable to the
halo crossing time of $\sim\! 80\Myr$ at $z \ssim 10$ (\equnp{tv} below).
The newly forming clouds of a given generation can be truly considered 
``feedback-free" only if they are shielded against feedback from star 
clusters of earlier generations that are in their active phase of feedback.
The main threat are expanding shock waves associated with 
stellar and supernova winds 
that can ablate or destroy the clouds in several cloud crushing times
via hydrodynamic instabilities (such as Kelvin-Helmholtz instability).
In order to determine which clouds survive and are capable
of FFB, we evaluate the cloud crushing time and compare it
to the free-fall time of $\sim\!1\Myr$.

\smallskip 
A conservative lower limit for the cloud
crushing time, derived in the absence of cooling and self-gravity,
is the time it should take the shock to cross the cloud, and is given by
\citep{klein94},
\be
\tcc = 2\,\frac{\Rc}{\Vw} \left(\frac{\rhoc}{\rhow}\right)^{1/2} \, .
\label{eq:tcc1}
\ee
Here, $\Rc$ and $\rhoc$ are the cloud radius and density,
and $\Vw$ and $\rhow$ are the wind velocity and density when it hits the cloud.
(The factor $2$ comes from the presence of a bow shock in the wind
outside the cloud,
which reduces the wind velocity and enhances the wind density by a
factor of $4$ each.)
We express $\Rc$ in terms of $\Mc$ and $\rhoc$.
In the absence of significant inter-stellar gas within the clusters that have
completed their FFBs 
\citep{menon23}, 
we can assume that the radiative energy losses are 
negligible, making the full supernova energy from the earlier clusters 
available for driving winds outside these clusters. 
We therefore assume that $\Vw$ is comparable to the original 
output wind velocity.
We write $\Vw \seq 3\stimes 10^3\kms\, \Vwthree$,
where the fiducial value is deduced from supernovae of mass $10\msun$ 
and energy $10^{51}\erg$ each.

once the clumps convert all the gas into stars, the radiative energy loss of the SN energy in the intra-clump medium is negligible. Therefore, the full SN energy  is available for driving winds at galaxy scale. 

\smallskip 
We assume that the feedback arises from an earlier generation of
clusters in their active-wind phase, of total mass
$\Mgen \seq 10^9\msun\,\Mgennine \seq 10^6\msun\,\Mgensix$.
These clusters are assumed to be distributed within a galaxy volume of radius
$\Rg \seq 1\kpc\, \Rgone$ (which we identify with the shock radius
in \se{nsh} and \se{shells}), 
namely at a characteristic distance $\sim\!\Rg$ from the forming cloud. 
The total mass flow rate at $\Rg$ is 
\be
\Mdotw = \Mgensix\, \dot{M}_{\rm w6}
  = 4\,\pi \Rg^2\,\rhow \Vw \, ,
\label{eq:Mdotw} 
\ee
where $\dot{M}_{\rm w6}$ is the mass-loss rate from each cluster of 
$10^6\msun$. Then
\be
\rhow = \frac{\Mdotw}{4\,\pi\,\Vw\,\Rg^2} \, .
\label{eq:rhow}
\ee

\smallskip
In order to evaluate $\dot{M}_{\rm w6}$, we used {\tt Starburst99} to compute 
the mass outflow rate from a burst in a cluster of $10^6\msun$, with different
IMFs and metallicities.
We learn from \fig{sb99_Mdot} that the peak of the wind is at a level of
\be
\dot{M}_{\rm w6} \simeq 0.01 \msun\,\yr^{-1}\, , 
\label{eq:Mdotw6}
\ee
and it lasts for $\tw \ssimeq 10\Myr$.
Inserting $\rhow$ from \equ{rhow} with \equ{Mdotw6} in \equ{tcc1} we obtain
\be
\tcc \simeq 5\Myr\, n_{3.5}^{1/6}\,\Mcsix^{1/3}\,
\Vwthree^{-1/2}\,\Rgone\,\Mgennine^{-1/2}\, .
\label{eq:tcc2}
\ee

\smallskip 
Simulations show that, with cooling and self-gravity ignored, the cloud
will mix into the ambient wind and be destroyed only
after several such cloud crushing times \citep{klein94}.
Cooling and self-gravity increase the survival time further, and may stabilize 
it forever if the cooling time in the turbulent mixing layer around the cloud 
and/or the free-fall time in the cloud are much shorter than $\tcc$
\citep{li20,gronke20,sparre20}.
Thus, as long as the free-fall time of $\sim\!1\Myr$ is shorter than $\tcc$, 
and so is the atomic cooling time of gas above $T\ssim 10^4$K, 
\equ{tcc2} provides a rather conservative lower
limit for the actual cloud crushing time.

\smallskip 
The mass of stars actively producing winds at a given time can be crudely
estimated from $\tw$.
For the total stellar mass of $\Ms \seq \eps\,\fb\,\Mv$,
and star formation that lasts for a halo virial time $\tv$ or more,
an upper limit for the fraction of stars in their active-wind phase 
is $\sim\!\tw/\tv$.
With $\tw \seq 10\Myr$ and $\tv$ from \equ{tv} below, this gives
\be
\Mgen \simeq  1.3\stimes 10^9\msun \,\eps\,\Mveight\,(1+z)_{10}^{3/2}\, ,
\label{eq:Mgen}
\ee
where $\Mv \seq 10^{10.8}\msun\, \Mveight$.
This is in the ball park of the fiducial value $\Mgennine \simeq 1$ 
used in \equ{tcc2}.
We will provide further estimates of $\Mgen$ in \se{clusters}.

\smallskip 
A cloud can be considered shielded against destructive winds
and capable of a FFB
once $\tcc$ is longer than the free-fall time of $\sim\!1\Myr$.
Using \equ{tcc2}, this implies a threshold cloud mass for shielding
of
\be
\Mcshield \simeq 0.8\stimes 10^4\msun\,n_{3.5}^{-1/2}\, \Vwthree^{3/2}\,
\Rgone^{-3}\, \Mgennine^{3/2}\, .
\label{eq:Mcshield}
\ee
We note that if $\tcc$ is compared to the general $\tff(n)$ instead of
$1\Myr$, the scaling of $\Mcshield$ with density becomes $\prop\!n_{3.5}^{-2}$.
If cooling and self-gravity increase $\tcc$ as estimated in \equ{tcc2}
by a factor of a few, say, the mass threshold would be smaller by this factor 
cubed.
We will show in \se{clusters} that the typical cloud masses are expected to be
$\Mc \ssim 10^6\msun$, namely they should be well shielded against winds from 
the early generations of stars.

\subsection{Shielding against UV radiation}
\label{sec:shield_uv}

Another worry, common at later times,
is shielding against ionizing UV radiation. At the relevant
$z \ssim 10$ regime, when the reionization and galaxy formation is
still limited to isolated bubbles, we do not expect a strong cosmological 
UV background.
However, as we did for the winds, we should estimate the potential damage
to a star-forming cloud from the radiation emitted by the earlier
generations of stars in the same galaxy.

\smallskip
The thickness $\Delta r$ of the UV shielding layer
can be estimated by equating the photon flux onto
the gas-cloud surface to the recombination rate per unit area.
The photon flux is
\be
\dot{Q} = \frac{f_{\rm OB}\,\nu_{\rm ion}\,\Mgen}{4\,\pi\,\Rg^2} \, ,
\ee
where
$\nu_{\rm ion} \seq 10^{49}\,{\rm s}^{-1}$ is the rate of ionizing photons
(over 13.6 eV) emitted by each O/B star,
and $f_{\rm OB} = 0.01\Msun^{-1}\,f_{{\rm OB},-2}$ is the number of OB
stars per unit mass.
The latter can be estimated by counting the stars more massive
than $20\msun$, the stars that produce H-ionizing photons 
\citep{sternberg03},
which for a Kroupa IMF is $f_{\rm OB} \ssimeq 0.005\msun^{-1}$.

\smallskip
On the other hand,
the recombination rate per unit area of the gas surface is
$n^2\,\alpha\,\Delta r$ per second,
where $n$ is the Hydrogen number density in the cloud,
and $\alpha \ssimeq 4\times10^{-13}\cm^3\,s^{-1}$
is the recombination rate per unit density of HII gas.

\smallskip
Equating these two rates provides the shielding length
\be
\Delta r = \frac{\nu_{\rm ion}\,f_{\rm OB}\,\Mgen}{4\,\pi\,\Rg^2\,n^2\,\alpha}
\sim 0.1\pc\, f_{{\rm OB},-2}\, \Mgennine\, \Rgone^{-2}\, n_{3.5}^{-2} \, .
\label{eq:Dr_UV}
\ee
This is negligible compared to the cloud radius of several parsecs,
as determined from Jeans or Toomre instability,
which implies that at these densities the clouds are well shielded also
against the UV radiation from earlier stars.

\smallskip 
Both \equ{tcc2} and \equ{Dr_UV} indicate that the assumption of earlier
clusters at a distance $d \ssim \Rg$ from the star-forming cloud did not 
lead to an overestimate of the cloud shielding.
The winds and radiation from each nearby cluster at $d \slt \Rg$
would indeed be more destructive, affecting $\tcc$ ($\Delta r$)
in proportion to $d$ ($d^{-2}$).
However, the mass in such clusters is $\prop\! d^3$ or $d^2$ 
(in the case of a planar disc), such that their contribution to
$\Mgen^{-1/2}$ ($\Mgen$) should balance the proximity effect.

\section{Cold inflow through the dark-matter halo}
\label{sec:cold_inflow}

Before we continue in \se{density} with the validity of the conditions for FFB 
in the galaxies at high redshift,
we make a detour to recollect the relevant properties of the host
dark-matter haloes as a function of mass and redshift, 
including the baryonic accretion onto them and through them.
We use these halo properties throughout, and in particular here 
to verify that the penetration of cold streams through the halo is 
efficient, recalling that efficient gas supply is a necessary condition 
for a globally efficient conversion of accreted gas to stars.

\subsection{Halo Properties}

\smallskip
According to the standard $\Lambda$CDM cosmology, with
current cosmological parameters $h=0.7$, $\omm=0.3$ and $\oml=0.7$,
the age of the Universe at $z$ is
\be 
t = 460\Myr\, (1+z)_{10}^{-3/2} \, ,
\ee 
where $(1+z) = 10\,(1+z)_{10}$.
For a halo of total virial mass  $\Mv$, 
the halo virial radius and velocity, $\Vv^2\seq G\,\Mv/\Rv$,
at a mean density contrast of $\Delta \seq 200$ above the cosmological 
background, are
\be
\Rv = 12.3\kpc\, \Mveight^{1/3}\, (1+z)_{10}^{-1} \, ,  
\label{eq:Rv}
\ee
\be
\Vv = 148\kms\, \Mveight^{1/3}\, (1+z)_{10}^{1/2} \, .  
\label{eq:Vv}
\ee
The fiducial value $\Mv \seq 10^{10.8}\msun$ is used throughout
(without loss of generality)
as it will turn out to be the threshold for $\eps \ssim 1$ at our fiducial 
redshift $1+z \ssim 10$, and it is the value deduced for the 
five brightest galaxies in the first {\tt CEERS} sample \citep{finkelstein23},
which are indeed near $z \ssim 9$ (see \se{obs}).
The halo viral crossing time is mass-independent at
\be 
\tv = \frac{\Rv}{\Vv} = 79\Myr\, (1+z)_{10}^{-3/2} \, .
\label{eq:tv}
\ee
For reference,
with the universal baryonic fraction of $\fb \seq 0.16$, 
the mean number density of baryons (using $\mu\seq 1.2$) is
\be
\nb = 2.2\times 10^{-4}\cmc\,(1+z)_{10}^3 \, ,  
\label{eq:nb}
\ee
and within the halo virial radius it is
\be
\nbv = 4.4\times 10^{-2}\cmc\,(1+z)_{10}^3 \, . 
\label{eq:nbv}
\ee 

\smallskip
The corresponding virial temperature, if there were a hot circum-galactic 
medium (CGM), is \be 
\Tv = 0.8\times 10^6 {\rm K}\, \Mveight^{2/3}\, (1+z)_{10} \, . 
\label{eq:Tv}
\ee
Assuming that the cold streams that feed high-redshift galaxies
\citep{bd03,keres05,db06,keres09,dekel09} 
flow in with a velocity comparable to the
virial velocity, and that their temperature is 
$T \ssim 10^4$K, corresponding to a sound speed of $c_{\rm s} \ssim 10\kms$,
the streams are supersonic with a Mach number 
\be
\Mach \ssim 15 \, \Mveight^{1/3}\, (1+z)_{10}^{1/2}\, T_4^{-1/2}\, .
\label{eq:Mach}
\ee

\smallskip
As shown analytically for an EdS cosmology
and confirmed by cosmological simulations \citep{dekel13}, 
the mean specific accretion rate onto a halo is
\be
\frac{\dot M}{M} \simeq 0.03\Gyr^{-1}\,M_{12}^{0.14}\,(1+z)^{5/2} \, .
\label{eq:sAR}
\ee
This is valid both for the total mass and for the baryons.
Using $\fb\seq 0.16$,
the mean baryonic accretion rate onto the halo is
\be
\Mdotac \simeq 65\msun\yr^{-1}\,\Mveight^{1.14}\,(1+z)_{10}^{5/2}\, .
\label{eq:Mdotac}
\ee
At $z \ssim 10$, this is about half the ``virial" accretion rate,
\be
\Mdotv = \frac{\fb\,\Mv}{\tv}
\simeq 128\msun\yr^{-1}\,\Mveight\,(1+z)_{10}^{3/2}\, .
\label{eq:Mdotv}
\ee
Thus, one may approximate $\Mdotac \ssimeq 0.5 \Mdotv$ at $z\ssim 10$,
when it is more convenient and when an accurate redshift dependence is 
not important.

\subsection{Cold inflow}

\smallskip 
In order to have all the accreted gas onto the halo turn into stars,
a complementary requirement to the feedback-free starbursts is that
most of the gas should penetrate cold from the halo virial
radius onto the central galaxy.
Efficient cold inflow is expected in the cold-flow regime,
especially when the hot component of the CGM is negligible.
Having only a little feedback-driven outflows in FFBs
is a necessary condition for no hot CGM.
Another necessary condition is having no stable virial shock that would have
heated the accreted gas to the virial temperature $\sim\!10^6$K 
\citep{bd03,db06}.
This is the case when the cooling time for a hypothetical shock-heated gas
in the outer halo is shorter than the time for compression 
behind the shock, which is comparable to the virial crossing time.  
A stable shock has been shown to form above a threshold halo 
mass of $\Mv \ssim 5\times 10^{11}\msun$, roughly independent of redshift in 
the range $z \seq 0 \sdash 5$ \citep{db06}. 
The exact mass depends on the cooling function 
at the assumed metallicity as a function of redshift \citep{sutherland93}.
Extending this calculation to higher redshifts, where the virial temperature
at a given mass is yet higher and it enters the Bremsstrahlung regime of the 
cooling function, 
yields a critical mass for shock heating of $\Mv \ssim 2\times 10^{12}\msun$ 
at $z \ssim 10$.
Our fiducial halo mass of $\Mv \ssim 10^{10.8}$ 
is thus well below the critical mass threshold for virial shock-heating, 
with $\tcool/\tvir \ssim 0.02$ in these haloes.

\smallskip
According to cosmological simulations,
after crossing the virial radius (where there is no stable shock), 
more than 90\% of the gas flows in freely through the halo 
via $\sim\!3$ dominant streams \citep{danovich12}. 
The atomic cooling time is
\be
\tcool = 1.3 \times 10^{-3}\Myr\, n^{-1}\, T_4\, \Lambda_{-22}^{-1}(T,Z) \, ,
\ee
where
$\Lambda=10^{-22}\, {\rm erg}\,{\rm cm}^{3}\,{\rm s}^{-1}\, \Lambda_{-22}$ 
is the cooling function, and $n$ is normalized to $1\cmc$.
The atomic cooling function for gas in collisional ionization equilibrium
\citep{sutherland93} has a local peak of $\Lambda_{-22} \ssimeq 1.3$ 
at $T \ssimeq 1.7\stimes 10^4$K, roughly independent of $Z$, 
from which it drops sharply toward lower temperatures to below 
$\Lambda_{-22} \ssim 10^{-3}$ just below $T \ssimeq 10^4$K.
At higher temperatures, $T \sgt 10^5$K, it can be crudely approximated by
\be
\Lambda_{-22} \simeq 0.09\,Z_{0.02}^{0.7}\,T_6^{-1} + 0.02\,T_6^{1/2} \, .
\label{eq:Lambda}
\ee 
Thus, during the inflow, the atomic cooling time is much shorter than the 
free-fall time of \equ{tff}, 
implying that the gas flows in roughly isothermally at $\gsim\! 10^4$K
until it hits the galaxy \citep[][Figs.~1 and 3]{bd03,db06}.
This puts these galaxies well in the cold-flow regime, where
the vast majority of the gas that is accreted onto the halo is expected to 
penetrate freely in streams at $T \ssim 10^4$K into the central galaxy 
on a virial timescale. No external mechanism that could heat the gas to higher
temperatures is expected at these redshifts.

\subsection{Stream radius}

\smallskip 
In certain calculations below we will need an estimate of the effective
cross section $\pi \Rs^2$ of the cold streams near the galaxy boundary 
$\Rg \ssim 1\kpc$, where $\Rs$ is the effective stream radius.
For three comparable streams this is roughly a factor
of $\sqrt{3}$ times the individual stream radius.
Since the haloes in question are high-sigma peaks (\se{density}), 
the stream width is expected to be significantly smaller than the halo 
radius \citep{db06,dekel09}.
The width of dark-matter filaments as estimated assuming top-hat 
cylindrical collapse into virial equilibrium
\citep[following][]{fillmore84} is
$\Rfilv \simeq \Rv\, (1+z)_{10}^{1/2}$ almost independent of halo mass.
The radius of the cold gas filaments has then been estimated 
\citep{mandelker18} assuming an angular-momentum conserving cylindrical 
contraction with a spin 
parameter $\lambda_{\rm s} = 0.025 \lambda_{{\rm s},.025}$.
This is analogous to the way the radius is estimated in spherical collapse.
The result for the effective radius of three filaments is
\be
\Rs \simeq 0.043\,\lambda_{{\rm s},.025}\, (1+z)_{10}^{1/2}\, \Rv \, .
\label{eq:Rs_mandelker}
\ee

\smallskip
Alternatively, it has been shown using a cosmological simulation of a 
Milky-Way-like galaxy at $z\seq 0$ \citep{ramsoy21} that for the dominant 
stream $\Rs/\Rv \ssim 0.2$ during the history of this galaxy. 
If extrapolated to $z \ssim 10$, the typical stream radius is
$\Rs \slsim 1\kpc$, where the halo mass is crudely expected to be 
$\sim\!10^{9}\msun$ \citep[][Fig.~1]{mandelker18}.
The mass dependence at a given $z$ is not clear. A lower limit for $\Rs$ in
an $\Mv \ssim 10^{10.8}\msun$ halo at $z\ssim 10$ could be obtained by 
assuming that $\Rs$ is the same at all masses, yielding a lower limit for 
the effective radius of three streams at
\be
\Rs \sim 0.043\, \Rv \, .
\label{eq:Rs_ramsoy}
\ee
This is in the same ball park as \equ{Rs_mandelker}.
We thus adopt $\Rs \ssimeq 0.7\kpc$ as our fiducial value at $1+z\ssim 10$.

\begin{figure} 
\centering
\includegraphics[width=0.49\textwidth,trim={1.5cm 5.5cm 1.5cm 4.6cm},clip]
{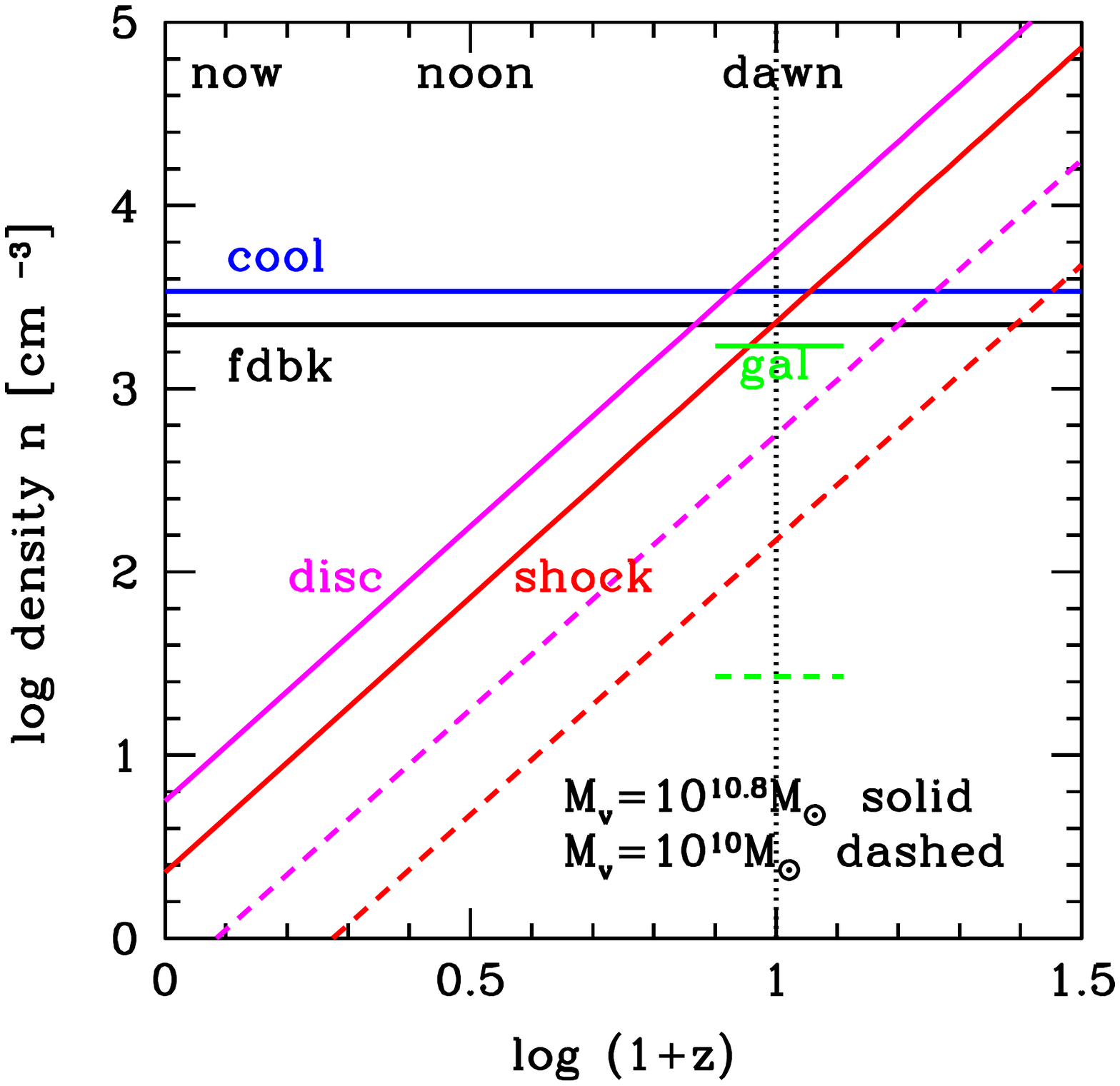} 
\vspace{-15pt}
\caption{
The gas densities $n$ as a function of redshift for two values of halo virial 
mass.
The feedback density $\nfbk$ (where $\tff \seq \tfbk$)
is from \equ{nfbk} (where $\tff \seq \tfbk$).
The cooling density $\ncool$ (where $\tcool \seq \tff$)
is from \equ{ncool} with $Z\seq 0.02$ and $C\seq 1$.
The post-shock density $\nsh$ is from \equ{nsh} with $\Rs\seq 0.7\kpc$.
The disc stellar density $\ndisc$ is from \equ{ndisc} with 
$\lambda\seq 0.025$ and $c\seq 1$.
The observed galaxy density $\nobs$ is from \equ{nobs} with 
$\Re \seq 0.3\kpc$ and $c\seq 1$.
The actual densities in the star-forming clouds are a factor $c \sgt 1$
larger than the galaxy averages shown.
Shown are the predictions for either $\Mv\seq 10^{10.8}\msun$ (solid) or
$\Mv\seq 10^{10}\msun$ (dashed).
In $\ndisc$ and $\nobs$ the assumed efficiencies are $\eps=1$ and $0.1$ for 
the two masses respectively.
We see for $\Mv\seq 10^{10.8}\msun$ that
at $z \ssim 10$ and above the densities $\ndisc$ and $\nsh$, as well as 
$\nobs$, are comparable to the critical densities $\nfbk$ and $\ncool$ 
for FFBs with $\eps\ssim 1$.
At lower redshifts, $\ndisc$, $\nsh$ and $\nobs$ are smaller than
the feedback-free critical densities, namely the feedback is effective causing
$\eps\ssim 0.1$.
At $z\ssim 10$, feedback-free starbursts are permitted only for
$\Mv \ssim 10^{10.8}\msun$ and above. 
They would be allowed also for lower masses
at higher redshifts, $z \ssim 20\sdash 30$.
}
\vspace{-10pt}
\label{fig:den}
\end{figure}

\section{Densities at cosmic dawn}
\label{sec:density}

Returning to the star formation in the galaxies, we next evaluate the density
expected in star-forming regions as a function of redshift and mass.
This will indicate when and where in the actual Universe
one expects densities as high as the critical density of $10^{3\sdash 4} \cmc$ 
for FFBs.
We first notice the density that can be potentially deduced directly 
from observations of the bright galaxies at $z\ssim 10$ for a given $\eps$.
We then estimate the expected densities in two different ways representing 
two limiting scenarios, as illustrated in \fig{shell}. 
One scenario is based on considering the shocks generated by the 
supersonic streams as they hit the galaxy boundary, assuming that they have a
substantial radial component.  
The other scenario considers galactic discs with star-forming 
clumps, as expected for streams that spiral-in with non-negligible
angular momentum and form Toomre-unstable discs.
These two scenarios will also be the basis for the analysis of star-forming
clusters in \se{clusters}.

At $z\ssim 10$, a halo of $10^{10.8}\msun$, being more massive by many orders
of magnitude than the Press-Schechter \citep{press74}
characteristic non-linear mass at that
time, emerged from a high-sigma density peak of about $5\sigma$.
Simulations indicate for such haloes that the feeding streams from the cosmic
web tend to be with lower angular momentum than for lower-sigma peaks, 
and thus more radial than at lower redshifts \citep{dubois12}.
The spin parameter for the incoming baryons may be $\lambda \ssim 0.025$ 
or even smaller.
This justifies the consideration of the two limiting scenarios.

\subsection{Density deduced from observations}
\label{sec:n_obs}

Given an initial gas mass that equals the 
stellar mass $\Ms\seq \eps\,\fb\,10^{10.8}\msun \, \Mveight$, 
and a gas effective radius comparable to the stellar effective radius 
$\Re = 0.3\kpc\,\Rethree$, 
where the fiducial values are motivated by the preliminary {\tt CEERS} 
observations at $z \ssim 10$ (\se{obs}),
the characteristic initial gas number density in the star-forming clouds is
\be 
\nobs \simeq \frac{c\,\Ms}{(4\pi/3)\Re^3}
     \simeq 1.7\times 10^3\cmc\, c\, \eps\, \Mveight\, \Rethree^{-3} \, .
\label{eq:nobs}
\ee
Here $c$ is the density contrast between the clouds and the mean
over the galaxy, which could be as large as $\sim \! 10$ (not to be confused
with the clumping factor $C$ in \equ{tcool}).
With $\eps \ssim 1$ and $c \sgsim 1$, 
the characteristic density indeed becomes $n \ssim 10^{3\sdash 4}\cmc$,
in the ball park of the critical density for FFBs.

\subsection{Post-shock density}
\label{sec:nsh}

The inflowing supersonic streams must come to a halt at a certain radius,
which can be associated with the galaxy boundary,
$\Rsh \ssim 1\kpc\,\Rshone$, where they pass through a strong shock.
Since the atomic cooling time at the relevant densities is extremely short, 
the shock would be isothermal, with the post-shock gas radiating its energy 
faster than waves can travel within the post-shock region. 
If the pre-shock density is $\rhoin$, the post-shock density for the isothermal
shock should be significantly larger,
\be
\rhosh = \rhoin\,\Mach^2 \, ,
\label{eq:nsh_Mach}
\ee
where the Mach number is $\Mach \ssim 15$ from \equ{Mach}.
The actual Mach number in the radial direction may be somewhat different;
larger because the stream accelerates on its way in,
or smaller because of an inclination angle between the stream and the shock 
front, which may be significant if the stream spirals in with high
angular momentum. We crudely assume hereafter that the relevant stream radial 
velocity is comparable to the virial velocity $\Vv$.

\smallskip 
The value of $\rhoin$ can be extracted from the baryonic mass inflow rate.
We assume that it is conserved between $\Rv$ and $\Rsh$,
where it is spread over a total stream cross section $\pi \Rs^2$,
namely
\be
\dot{M}_{\rm ac} = \pi\, \Rs^2\, \rhoin\, \Vv\ \, .
\label{eq:rhoin}
\ee
Based on \equ{Rs_mandelker} and \equ{Rs_ramsoy},
we express $\Rs = 0.7 \kpc\, \Rsseven$. 
Using the accretion rate from \equ{Mdotac}, we obtain
\be
\nin \simeq 10 \cmc\,\Rsseven^{-2}\, \Mveight^{0.81}\, (1+z)_{10}^2\, .
\label{eq:nin}
\ee

\smallskip 
Substituting $\nin$ from \equ{nin} and $\Mach$ from \equ{Mach} in
\equ{nsh_Mach}, we obtain for the post-shock shell density 
\be
\nsh \simeq 2.3 \times 10^3\cmc\, c\,           
\Rsseven^{-2}\, \Mveight^{1.48}\,T_4^{-1}\, (1+z)_{10}^{3}\, .
\label{eq:nsh}
\ee
Here $c$ is the density contrast between the star-forming clouds and
the mean post-shock shell density, which grows above unity when the clouds
collapse.

\smallskip 
Despite the uncertainties, we conclude that in haloes of
$\Mv \ssim 10^{10.8}\msun$ at $z \ssim 10$
the post-shock density is expected to be in the
ball park of $10^{3\sdash 4}\cmc$.
As shown in \fig{den}, 
this is comparable to the densities for FFBs as derived
in \se{fbk} and \se{burst}.
Within the first stages of fragmentation to clouds (\se{clusters}), the density
will cross the cooling threshold and allow FFBs 
that will lead to $\eps \ssim 1$.

\subsection{Disc clump density}
\label{sec:disc}

The mean baryon density in high-$z$ discs can be estimated from the mean
baryon density within the halo, $\nbv$ of \equ{nbv}, 
$\ndisc \simeq 2\,c\,\lambda^{-3} \, \nbv$.
The factor $\lambda$ represents the contraction from the halo virial radius 
to the effective galaxy stellar radius, $\lambda \seq \Re/\Rv$.
If the galaxy is a thick disc of axial ratio $\Rd/\Hd$, 
there is an additional geometrical factor of $(2/3)(\Rd/\Hd)$,
which is $\sim\!2$ for $\Rd/\Hd \ssim V/\sigma \seq 3$.
The factor $c \sgt 1$ is the density contrast between the star-forming 
clumps and the mean background density in the disc.
This gives
\be
\ndisc \simeq 
5.6\times 10^3\cmc\, c\, \lambda_{.025}^{-3}\, (1+z)_{10}^3 \, . 
\label{eq:ndisc}
\ee
When referring to the stellar density, $\ndisc$ should be multiplied by $\eps$.

\smallskip 
A value of $\lambda \sim 0.03$ is valid in typical observed and simulated
galaxies at lower redshifts \citep{somerville18,jiang19}.
At $z\ssim 2$, it relates a halo of $\Mv \ssim 10^{12}\msun$ with 
$\Rv \ssim 100\kpc$ to a galaxy with $\Re \ssim 3\kpc$, 
typical values for massive star-forming discs \citep{burkert16}.
The contraction factor $\lambda$ is sometimes identified with the universal 
halo spin parameter with conservation of angular momentum 
\citep{fall80,danovich15}.
For the high-sigma-peak galaxies of $\Mv \seq 10^{10.8}\msun$ at $z \ssim 10$,
with $\Rv \ssim 12.3\kpc$ from \equ{Rv}, a spin parameter of
$\lambda \ssim 0.025$ leads to $\Re \ssim 0.3\kpc$. 
A merger-driven compaction process may lead to even smaller values of 
$\lambda$ \citep{zolotov15,lapiner23}.
These values of $\Re$ are consistent with the observed values in the first
CEERS sample, where the median is $\Re \ssim 0.3\sdash 0.4\kpc$ (\se{obs}).
As shown in \fig{den}, at $z \ssim 10$ and $\Mv \ssim 10^{10.8}\msun$,
the disc gas density in \equ{ndisc} is comparable to the post-shock density
of \equ{nsh}, and it is consistent with the empirical stellar 
density of \equ{nobs} for $\eps \ssim 1$.
As we will see in \se{clusters},
disc fragmentation to clumps will increase $c$ in the clumps
by an order of magnitude \citep{ceverino12,mandelker14}.
This will make the density in the star-forming clouds significantly higher 
than the thresholds for FFBs, \equ{nfbk} and \equ{ncool}.

\begin{figure*} 
\centering
\includegraphics[width=0.80\textwidth,trim={4.0cm 3.5cm 3.8cm 2.5cm},clip]
{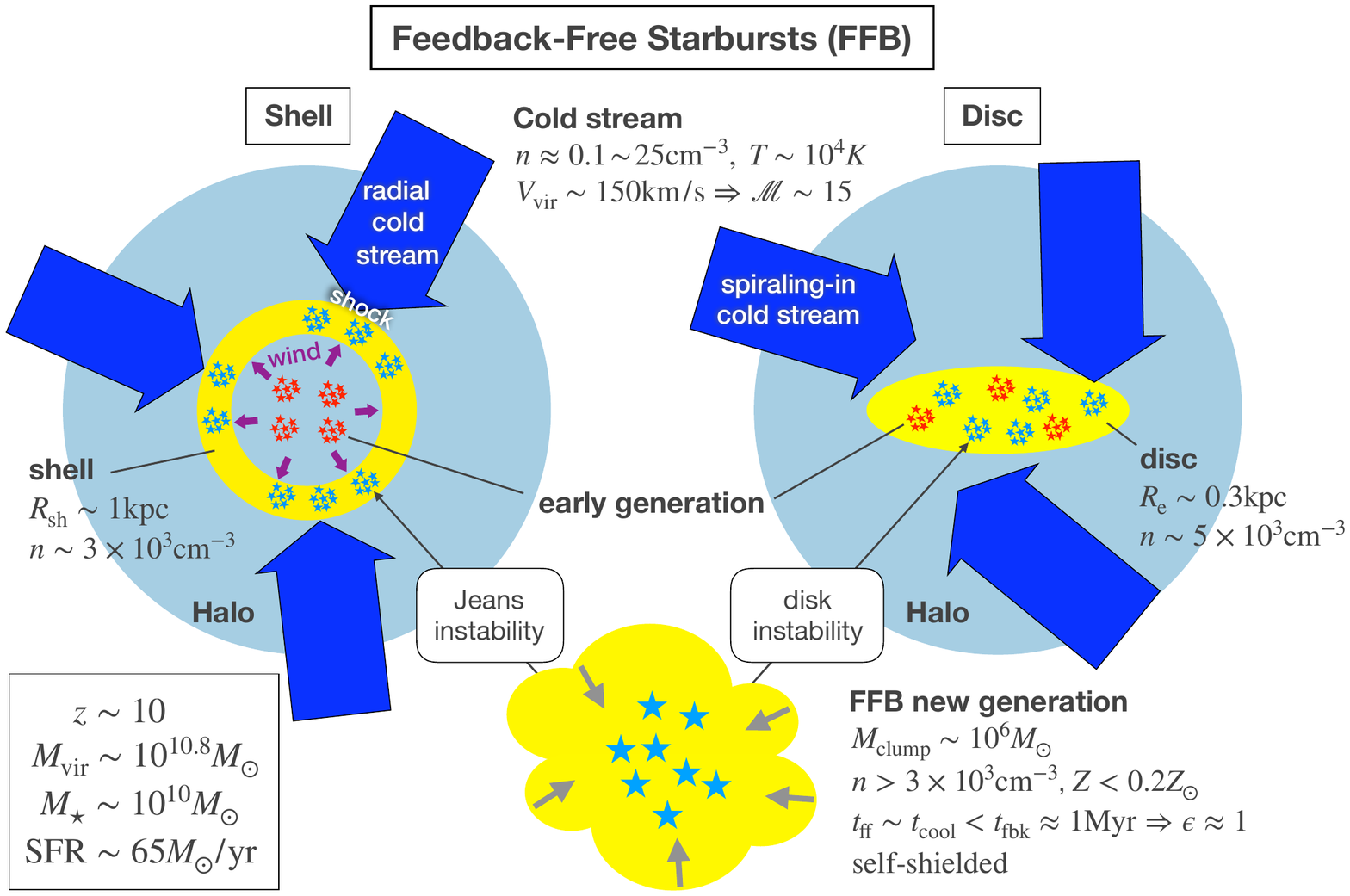} 
\caption{
A cartoon illustrating a generation of feedback-free starbursts
in massive clusters within a shell or a disc of
$n \ssim 3\times 10^3\cmc$ at $r \slsim 1\kpc$.
The shell is confined by the shock generated by the inflowing supersonic cold
streams and by the wind from an earlier generation of stars, while
the disc size is determined by angular momentum.
Shell fragmentation is allowed after the gas accreted locally exceeds the
Jeans mass, while the disc fragments after the accreted gas brings the Toomre
$Q$ to below unity.
Starbursts follow in the clusters as $\tcool \slt \tff$.
}
\vspace{-10pt}
\label{fig:shell}
\end{figure*}

\section{Star clusters and generations}
\label{sec:clusters}

The starbursts are predicted to occur in clouds that fragment from the gas 
in the galaxy as it is gradually fed by the cold streams.
We address the expected cloud properties in the two scenarios, of spherical 
shells and of discs, as illustrated in \fig{shell}. 
In each scenario, the gas is accumulating until the accreted gas mass grows 
above a threshold that allows fragmentation into the clouds.
We show that this leads to several generations of star-cluster formation.

\subsection{Jeans clouds in shells}
\label{sec:jeans}

\subsubsection{Jeans radius and mass}

In the scenario where the streams flow in rather radially, we
expect the fragmentation to starbursting clouds to occur in parts of a shell of
radius $\Rsh \slsim 1\kpc$ in which the density is $\nsh \ssim 10^{3.5}\msun$
as predicted in \se{nsh} and discussed in more detail in \se{shells}.
Given the density and temperature of the gas in this shell, 
the Jeans radius (half the Jeans wavelength) 
for a self-gravitating gas with a sound speed $\cs$ corresponding to $T$ is
\be
\Rj = \left( \frac{\pi \cs^2}{4 G\rho} \right)^{1/2}
\simeq 14.4\pc\, T_4^{1/2}\, n_{3.5}^{-1/2} \, .   
\label{eq:Rj}
\ee
The associated Jeans mass is
\be
\Mj \simeq 1.06 \times 10^6\msun\, T_4^{3/2}\, n_{3.5}^{-1/2} \, , 
\label{eq:Mj}
\ee
above which perturbations are Jeans unstable \citep{jeans02}.
In fact,
an isothermal self-gravitating cloud in a pressure confined medium, 
such as our shell, can only retain hydrostatic equilibrium and avoid
gravitational collapse if its mass is below the Bonnor-Ebert (BE) mass 
\citep{bonnor56}.
For a given temperature of the cloud, due to pressure continuity 
at the cloud-medium boundary, the pressure dependence of the BE mass 
can be converted to a dependence on the density in the cloud.
It turns out that the value of the BE mass is very similar to the value of the 
Jeans mass, and they serve a similar purpose, despite the fact that they are 
derived in different ways.
The fragments are thus expected to be Jeans or Bonnor-Ebert clouds
that resemble massive proto-globular clusters.
Non-linear fragmentation may eventually lead to smaller sub-clouds.

\smallskip
Based on \equ{Mcshield} and \equ{Dr_UV}, clumps of such mass and density
are expected to be shielded against winds and radiation from earlier 
generations of star clusters, and thus capable of forming FFBs.

\subsubsection{Generations in shell clouds}

For the shell to be able to fragment by 3D Jeans instability, 
when the cooling time is longer than the free-fall time, 
the accreting gas should accumulate until the mass within a Jeans area on 
the shell becomes comparable to the Jeans mass.
This defines a generation of Jeans clouds within which the density could grow
above the critical cooling density for star formation.
After accretion of mass $\Mgen$ trough streams of total area $\pi\Rs^2$, 
where the Jeans area is $\pi\Rj^2$, this condition is
\be
\Mgen \frac{\Rj^2}{\Rs^2} \simeq \Mj \, .
\ee
Inserting $\Mj$ and $\Rj$ from \equ{Mj} and \equ{Rj}, the mass involved in each
generation is
\be
\Mgen \simeq 2.5\times 10^9\msun\, \Rsseven^2\, T_4^{1/2}\, n_{3.5}^{1/2} \, . 
\label{eq:Mgen_Jeans}
\ee
This mass, being somewhat higher than our fiducial value of 
$\Mgen \ssim 10^9\msun$,
would raise the threshold cloud mass for shielding as estimated in
\equ{Mcshield} by a factor of $\sim\!4$,
but $\Mc \ssim 10^6\msun$ clouds will still be safely self-shielded.

\subsection{Disc clumps}
\label{sec:disc_clumps}

\subsubsection{Disc Clump Mass and radius}
\label{sec:disc_clump_mass}

In the scenario where the streams spiral in with non-negligible angular 
momentum and settle to a galactic disc, the disc will fragment once the 
Toomre $Q$ parameter \citep{toomre64} is smaller than unity.\footnote{A word of
caution here is that fragmentation may also occur in $Q \sgt 1$ situations 
where there are excessive compressive modes of turbulence \citep{inoue16}.}
This parameter can be expressed as \citep{dsc09,ceverino12}
\be
Q \simeq \frac{\sigd\,\kappa}{\pi\,G\,\Sigd}
\simeq \sqrt{2}\,\delta^{-1}\, \frac{\sigd}{\Vd}\, .
\label{eq:Q}
\ee
Here,
$\sigd$ is the radial velocity dispersion in the disc,
$\Vd$ is the rotation velocity, 
$\kappa$ is the associated epicyclic frequency that relates to the angular 
velocity $\Omega \seq \Vd/\Rd$ as $\kappa \seq \sqrt{2}\,\Omega$ for a flat 
rotation curve, 
and $\Sigd$ is the surface density of the cold disc component, 
which we crudely identify below with the disc gas of mass $\Md$ and radius 
$\Rd$. 
The second equality refers to the key variable $\delta$ \citep{dsc09},
the ratio of cold disc mass to the total mass encompassed within $\Rd$,
including baryons and dark matter, 
\be
\delta = \frac{\Md}{\Mtot} \, .
\label{eq:delta}
\ee
The pre-collapse radius of the clumps that forms under Toomre
instability, representing the fastest growing mode, is then
\be
\RT = \frac{\pi}{4}\, Q\,  \delta\, \Rd \, .
\label{eq:RT}
\ee
Assuming that the surface density within the clumps follows the disc surface
density, the corresponding Toomre mass is 
\be
\MT \simeq \frac{\pi^2}{16}\,Q^2\,\delta^2\,\Md \, .
\label{eq:MT}
\ee

\smallskip 
Once $Q \sleq 0.67$, the thick disc is unstable \citep{goldreich65_thick},
and in practice it fragments to clumps with masses comparable to $\MT$ 
and lower.  
For massive discs at high redshifts, 
we expect thick discs in which the instability is
regulated by inward radial transport and accretion, with 
$\delta \ssim 0.3$ \citep{dsc09,ginzburg22}. 
An upper limit for the Toomre mass can be obtained by assuming that 
the cold disc is made of all the baryons ever accreted, 
$\Md \ssim \fb \Mv$ in \equ{MT}, which gives 
\be 
\MT \simeq 2.5\times 10^8\Msun\, Q_{0.67}^2\,\delta_{0.3}^2\,\Mveight\, ,
\label{eq:MT1}
\ee
where $Q \seq 0.67\,Q_{0.67}$ and $\delta \seq 0.3\,\delta_{0.3}$.

\smallskip 
For an alternative  estimate of the Toomre mass,
we assume that at the relevant high redshifts
the disc turbulence, which otherwise tends to decay on a disc 
dynamical timescale $\td$, is being driven efficiently by the 
accretion \citep{ginzburg22}, converting most of the stream kinetic energy to
turbulence, namely
\be
\frac{3}{2}\,\Md\,\sigd^2 \simeq \frac{1}{2}\,\Mdotac\, \td\, \Vv^2\, .
\label{eq:turb}
\ee
Here, $\Md$ is the gas mass per generation, to be derived in \se{generations},
and the stream inward velocity is assumed to be $\Vv$.
We next assume that the disc rotation velocity is
\be
\Vd^2 \simeq \frac{\fb}{2\lambda}\,\Vv^2 \, .
\label{eq:Vd}
\ee
This is based on $\Vd^2 \seq G\,\Mt/\Rd$ while $\Vv^2 \seq G\,\Mv/\Rv$,
with the definition $\Rd \seq \lambda\Rv$ for the half-mass radius,
and the crude approximation, assuming $\epsilon\simeq 1$,
that the total mass interior to $\Rd$ is
\be
\Mt \simeq 0.5\,\fb\,\Mv \, . 
\label{eq:Mt}
\ee
Combining \equ{Vd} with \equ{turb} and \equ{Q}, we obtain 
\be
\Md \simeq \frac{4}{3}\, \frac{\lambda}{\fb}\, \Mdotac\,\td\, 
Q^{-2}\,\delta^{-2} \, .
\label{eq:Md1}
\ee
Inserting $\Md$ in \equ{MT}, we obtain
\be
\MT \simeq \frac{\pi^2}{12}\, \frac{\lambda}{\fb}\, \Mdotac\,\td \, .
\label{eq:MT2}
\ee
Interestingly, the explicit dependence of $\MT$ on $Q$ and $\delta$ from 
\equ{MT} disappears once the turbulence is assumed to be driven
by accretion, \equ{turb}.

\smallskip
Adopting for $\Mdotac$ the average baryonic accretion rate from \equ{Mdotac}
and using $\td \seq \Rd/\Vd$ and $\lambda \seq \Rd/\Rv$,
we obtain for the mass that is accreted in a disc dynamical time
\be
\Mdotac\,\td \simeq 7.2 \times 10^7 \Msun \,           
\lambda_{.025}^{3/2}\,\Mveight^{1.14}\,(1+z)_{10} \, .
\label{eq:Mdotac_td}
\ee
Inserting this in \equ{MT2}, with $\fb \seq 0.16$, we finally get
\be
\MT \simeq 9.2\times 10^6\msun\,\lambda_{.025}^{5/2}\, \Mveight^{1.14}\, 
(1+z)_{10} \, .           
\label{eq:MT3}
\ee
We learn that the Toomre mass is not only increasing with halo mass, 
as expected, but is also increasing with redshift, due to the decrease of the 
accretion time with respect to the dynamical time. 
We note that the Toomre mass that is relevant for clumps in the disc scenario
is somewhat larger than the Jeans mass that is relevant
in the shell scenario, \equ{Mj}.
We expect $\MT$ to be an upper limit for the range of clump masses in the 
discs. 

\subsubsection{Gas consumption in clumps}

\smallskip 
A necessary condition for the FFB scenario is that most of the gas is consumed
into stars in star clusters.
In discs at the moderate redshifts of cosmic noon, 
clumps below $\sim\!10^8\msun$ tend to disrupt
by feedback from their own stars
\citep{mandelker14,mandelker17,dekel23_gclumps},
thus leading to only partial gas consumption into long-lived star-forming 
clumps.
In the $z \ssim 10$ discs, the gas consumption can be significantly higher
because of the feedback-free starbursting.
Since the Toomre mass is well above the threshold of
$\Mcshield \ssim 10^4\msun$ for shielding against feedback from
other clusters (\equnp{Mcshield}),
we expect a large fraction of the gas to be in shielded clumps.
To estimate this fraction, we recall that, based on zoom-in cosmological
simulations of discs \citep{mandelker14,mandelker17},
the non-linear clump mass function is expected to be close to a
scale-free power-law
\be
\frac{dN}{dm} \prop m^{-\alpha} \, ,
\label{eq:mass_func}
\ee
with $\alpha \slsim 2$. It extends from near $\MT$
down to below the resolution scale of the simulation.
This kind of mass function is a generic result in a supersonic turbulent medium
\citep{hopkins13_frag,trujillo19,gronke22}.
Assuming, for example, $\alpha \seq 1.8$ \citep{mandelker17},
with $\MT$ from \equ{MT3} and $\Mcshield$ from \equ{Mcshield},
one obtains that the fraction of disc mass in shielded clumps is
larger than $1 - (\Mcshield/\MT)^{2-\alpha}$, namely $79\%$.
This allows most of the gas to participate in the feedback-free star
formation, thus permitting an overall $\eps \ssim 1$.

\subsubsection{Generations in disc buildup}
\label{sec:generations}

\smallskip
A necessary condition for starbursts in the disc scenario
is the very formation of clumps, namely disc instability with 
$Q \sleq 0.67$. This introduces yet another threshold
which divides the overall star formation in the disc galaxy into several
generations of starbursts.
For a crude estimate, we assume that the disc gas is largely depleted during 
each generation of starbursts, as argued in the preceding paragraph.
We also assume for simplicity that the cold disc mass $\Md$ that drives the 
instability is dominated by the gas mass, with most of the stars kinematically 
hotter.
These two assumptions imply that after the end of each generation 
the value of $\Md$ is low, such that $\delta \sll 1$, and therefore
the disc is stable with $Q\sgg 1$.
As the gas mass is gradually building up again by accretion,
$\Md$ is growing, until $Q$ becomes smaller than the critical value of $0.67$ 
and the disc becomes unstable. This is the onset of a new generation of
starbursts, and so forth.

\smallskip 
In order to estimate the mass involved in each generation, 
we express $\Md$ by \equ{Md1}, with $\Mdotac\,\td$ based on \equ{Mdotac_td}.
Using the definition of $\delta$ in \equ{delta}, 
with the assumed $\Mt$ from \equ{Mt}, 
we have $\delta^{-2} \ssimeq 0.25\, \fb^2 \Mv^2 /\Md^2$. 
Solving for $\Md$, we obtain 
for the disc mass that is built in each generation  
\be
M_{\rm gen} = \Md \simeq 0.95\times 10^9\msun\,  
Q_{0.67}^{-2/3}\,\lambda_{.025}^{5/6}\,\Mveight^{1.05}\,(1+z)_{10}^{1/3} \, .
\label{eq:Mgen_disc}
\ee
This implies more than ten generations of FFB starbursts, 
comparable and somewhat larger than the number of generations 
deduced in the shell scenario, \equ{Mgen_Jeans}.

\section{Redshift and mass thresholds for FFB}
\label{M_z}

We are now in a position to summarize the predicted thresholds in redshift and
mass for FFB galaxies in the two scenarios of shells and discs.

\subsection{FFB in shells}

\begin{figure} 
\centering
\includegraphics[width=0.49\textwidth,trim={1.5cm 5.5cm 1.5cm 4.6cm},clip]
{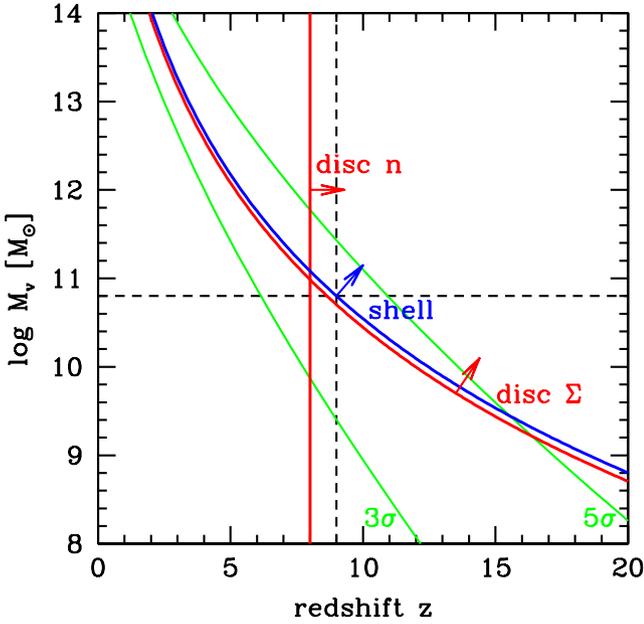}
\vspace{-15pt}
\caption{
Thresholds for FFB with $\epsilon \ssim 1$ in the redshift and halo-mass plane.
For the shell scenario, the blue line marked ``shell" shows the threshold based 
on \equ{M_z_shell}, which is relevant both for the 3D and 2D density 
criteria.
For the disc scenario, the vertical red line marked ``disc $n$"
shows the redshift threshold based on the 3D density criterion from
\equ{z_ffb_disc} (with $c\seq 8$).
The mass dependence in the disc case enters in the diagonal red line marked 
``disc $\Sigma$", which shows the complementary threshold based on the
surface-density criterion from \equ{M_z_disc} 
(with $\Sigma_{\rm crit}\seq 2\times 10^3 \msun\,\pc^{-2}$).
The green curves mark the corresponding $3\sigma$ and $5\sigma$ peaks in the
Gaussian density fluctuation field. 
We learn that at $z \ssimeq 9$, the threshold mass for FFB lies near 
$\Mv \ssimeq 10^{10.8}\msun$. 
At higher redshifts, the mass threshold becomes less demanding,
steeply dropping to $\Mv \ssim 3.5\times 10^9\msun$ at $z \seq 15$
and to $\Mv \ssim 6\times 10^8\msun$ at $z \seq 20$.
It roughly follows the $4\sigma$ peak mass near $z \ssim 10$, but the critical
haloes gradually become more rare at higher redshifts.
On the other hand, FFB is less likely at $z \slt 8$, where it is ruled out in 
the disc scenario and is limited to the rare massive haloes in the 
shell scenario. 
}
\vspace{-10pt}
\label{fig:mz}
\end{figure}

We note that $\nsh$ in \equ{nsh} depends on both redshift and mass.
Assuming $\Rs \ssim 0.05\,\Rv$ based on the lower estimate \equ{Rs_ramsoy},
with $\Rv$ from \equ{Rv}, we obtain in \equ{nsh}
\be
\nsh \sim 3\times 10^3 \cmc\,\Mveight^{0.81} (1+z)_{10}^5 \, .
\ee
Then, the basic FFB requirement on the 3D density,
$\nsh \sgt \nfbk$ from \equ{nfbk}, yields a necessary condition for FFB
in the $\Mv \sdash z$ plane at 
\be
\Mveight\, (1+z)_{10}^{6.2} > 1 \, .
\label{eq:M_z_shell}
\ee
This threshold is shown as the blue line marked ``shell" in \fig{mz}.

\smallskip 
For the complementary necessary condition for FFB based on the clump
surface density, $\Sigma \sgt \Sigma_{\rm crit}$ of \equ{Sigma_crit},
required to make the radiative feedback ineffective,
we adopt the cloud Jeans radius $\Rj \sprop n^{-1/2}$ from \equ{Rj}.  
Using the relation between 2D and 3D density in the cloud via the clump radius,
\equ{Sigma_n}, the crude condition on the surface density becomes
$n_{3.5}^{1/2} \sgt 1$.
This condition turns out to coincide with the condition implied by the basic 
requirement on the 3D density $n \sgt \nfbk$, namely \equ{M_z_shell}.

\subsection{FFB in discs}

We learn from the overall disc density in \equ{ndisc} that the basic condition 
for FFB in a disc, $\ndisc \sgt \nfbk$, translates to a threshold in redshift
with no explicit mass dependence.
This is as opposed to the strong mass dependence of $\nsh$ in \equ{nsh},
which translates to a strong redshift dependence of the threshold mass
in \equ{M_z_shell}.
For an estimate of the 3D gas density in the disc in each generation,
we use the disc mass from \equ{Mgen_disc}, with 
$\Hd = 0.33\,\Rd\, (H/R)_{0.33}$, to obtain
\be
\ndisc \simeq 0.58 \times 10^3\cmc\, c\, 
Q_{0.67}^{-2/3}\, \lambda_{.025}^{-13/6}\, \Mveight^{0.05}\, (1+z)_{10}^{10/3} 
\, ,
\label{eq:nd_gen}
\ee
which also scales with $(\Hd/\Rd)_{0.33}^{-1}$.
The mass dependence is indeed negligible.
Requiring $\ndisc \sgt \nfbk$ from \equ{nfbk}, 
we obtain a pure redshift threshold for FFB near
\be
(1+z)_{10} > 1.5\, c^{-0.3}\, Q_{0.67}^{0.2}\, \lambda_{.025}^{0.65} \, , 
\label{eq:z_ffb_disc}
\ee
which also scales with $(\Hd/\Rd)_{0.33}^{0.3}$.
For the fiducial choice of parameters, and with $c \ssim 8$, say,
the crude condition is $1+z \sgt 8$. 
This is shown as the vertical red line in \fig{mz}.

\smallskip 
However, a mass dependence enters also in the disc scenario via the 
dependence of the complementary surface-density threshold for FFB,
\equ{Sigma_crit}, on the radius of the star-forming clumps.
As seen in \equ{Sigma_n},
given the FFB 3D density $n$, the validity of the surface density criterion
depends on the clump radius $\Rc \seq c^{-1} \RT$.
From \equ{Mgen_disc} and the assumed $\Mt$ from \equ{Mt} we get
\be
\delta \simeq 0.19\,  Q_{0.67}^{-2/3}\,\lambda_{,025}^{5/6}\, \Mveight^{0.05}\,
(1+z)_{10}^{1/3} \, .  
\ee
Inserting this in \equ{RT}, with $\Rd \seq \lambda\,\Rv$ and
$\Rv$ from \equ{Rv}, we obtain
\be
\Rc \simeq 30\pc\,c^{-1}\, Q_{0.67}^{1/3}\, \lambda_{.025}^{11/6}\,
\Mveight^{0.38}\,(1+z)_{10}^{-2/3} \,.  
\label{eq:Rc}
\ee
Using $\Rc$ from \equ{Rc} in \equ{Sigma_n}, with $\ndisc$ from \equ{nd_gen},
we obtain that the approximate condition on surface density, 
$\Sigma \sgt \Sigma_{\rm crit}$ from \equ{Sigma_crit}, becomes in the case of 
disc clumps
\be
\Mveight\,(1+z)_{10}^{6.2} > 2.0\, 
Q_{0.67}^{0.77}\, \lambda_{.025}^{0.77}\, \Sigma_{{\rm crit},3.5}^{2.33}\, .
\label{eq:M_z_disc}
\ee
This scales with $(\Hd/\Rd)_{0.33}^{2.32}$.
With the fiducial parameters, 
and with $\Sigma_{\rm crit} \ssim 2\times 10^3\msun\pc^{-2}$,
the surface-density condition for FFB becomes 
$\Mveight\,(1+z)_{10}^{6.2} > 0.8$,
which is shown as the diagonal red line in \fig{mz}.
We learn that this threshold for FFB in the disc scenario
is surprisingly similar to the corresponding threshold in the shell scenario,
\equ{M_z_shell}, shown as the blue line in \fig{mz}.

\smallskip
Thus, \equ{M_z_shell} and \equ{M_z_disc} provide a robust condition for
FFB in the mass-redshift plane.
The threshold mass for FFB, which is $\Mv \ssimeq 10^{10.8}\msun$ 
at $z \seq 9$, drops steeply at higher redshifts,
reaching $\Mv \ssim 3.5\times 10^9\msun$ at $z \seq 15$
and $\Mv \ssim 6\times 10^8\msun$ at $z \seq 20$.
We learn that FFB is unlikely at $z \slt 8$, where it is excluded in
the case of a disc and is limited to increasingly rare massive haloes in the
case of a shell.

\section{Shells of starbursting clouds}
\label{sec:shells}

In the scenario where the inflowing streams are rather radial we expect the
starbursting clouds of each generation to form in a thin shell of radius 
$\Rsh$, as illustrated in \fig{shell}.
The shell radius, which we tentatively assumed to be on the order of $1\kpc$,
enters several of the estimates performed above and determines the final size
of the galaxy.
We already estimated in \equ{nsh} that the post-shock density of the
instreaming gas is expected to be on the order of the characteristic density 
of a few $\times 10^3\cmc$.
The stellar winds and supernova remnants of earlier generations are expected
to join to a hot bubble that confines the shell from the inside. 
We evaluate the shell stalling radius by equating the interior ram 
pressure of the bubble and the ram pressure of the instreaming gas.
%
The pressure balance reads
\be
\rhow \Vw^2 = \rhoin \Vv^2 \, ,
\label{eq:ram}
\ee
assuming that the stream velocity is $\Vv$.
Considering a shell of radius $\Rsh$,
the l.h.s can be read from \equ{Mdotw} with $\Rg \seq \Rsh$, namely
$\rhow \Vw^2 \seq \Mdotw \Vw / (4\pi\Rsh^2)$.
Considering streams with an effective radius $\Rs$,
the pre-shock density $\rhoin$ as deduced from \equ{rhoin} gives in the r.h.s
$\rhoin \Vv^2 \seq \Mdotac \Vv / (\pi \Rs^2)$.
Inserting these in \equ{ram} we obtain for the shell radius
\be
\frac{\Rsh^2}{\Rs^2} = \frac{\Vw \Mdotw}{4 \Vv \Mdotac} \, .
\label{eq:Rsh_ram1}
\ee
With $\Mdotac$ from \equ{Mdotac} and $\Vv$ from \equ{Vv},
and with $\Mdotw$ from \equ{Mdotw} plus \equ{Mdotw6}
and a free supernova wind of $\Vw \seq 3\times 10^3\kms$,
we obtain
\be
\frac{\Rsh}{\Rs} = 1.25\, \Mveight^{-0.73} \, (1+z)_{10}^{-3/2}\,
\Mgennine^{1/2}\, \Vwthree^{1/2} \, .  
\label{eq:Rsh_ram}
\ee
With $\Rs \ssim 0.7\kpc$ (\equnp{Rs_mandelker} and \equnp{Rs_ramsoy}),
we conclude that $\Rsh \slsim 1\kpc$.
The above crude estimate justifies the fiducial value of $\Rsh \slsim 1\kpc$
adopted in our analysis and our prediction of compact FFB galaxies of sub-kpc 
radii.

\section{Comparison to JWST observations}
\label{sec:obs}

\begin{figure} 
\centering
\includegraphics[width=0.49\textwidth,trim={0.5cm 0.75cm 0.4cm 0.80cm},clip]
{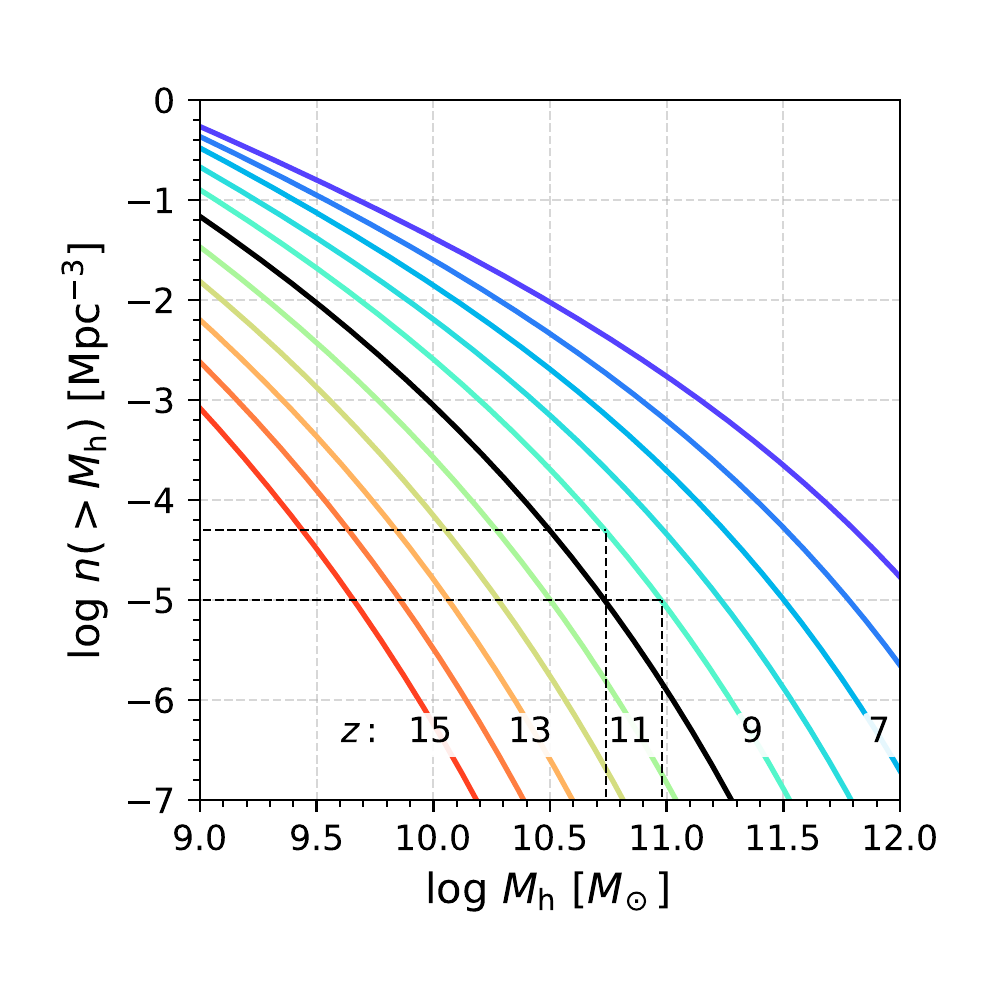} 
\vspace{-15pt}
\caption{
The cumulative halo mass function at different redshifts \citep{watson13}.
It assumes a $\Lambda$CDM cosmology with 
$\omm\seq 0.3$, $\oml\seq 0.7$, $h\seq 0.7$ and $\sigma_8\seq 0.82$. 
The haloes are spherical with a mean density contrast of 200.
This serves for estimating the halo masses of galaxies 
in a complete sample within a given effective volume, 
assuming rank-preserving abundance matching between the galaxy 
luminosities and the halo masses. 
With a volume of $10^5\Mpc^3$, the halo mass of the brightest galaxy at 
$z \ssimeq 9$ is expected to be $\Mv \ssimeq 10^{11}\msun$,
and the 5th brightest galaxy is expected to have $\Mv \ssim 10^{10.8}\msun$,
as marked by the dashed lines.
}
\vspace{-10pt}
\label{fig:hmf}
\end{figure}

The first samples of galaxies that were detected by {\tt NIRCam} on {\tt JWST}
with photometric redshifts  $z\seq 7\sdash 15$ 
\citep{naidu22,finkelstein22b,finkelstein23,donnan23a,donnan23b,labbe23,
robertson23,tacchella23a,tacchella23b}
are consistent with the main robust prediction of our analysis;
a high efficiency $\eps$ of conversion of accreted gas into stars
in the $z \ssim 10$ massive dark-matter haloes with densities of a few
$\stimes 10^3\cmc$. 
This is translated to an excess of
bright galaxies compared to the predictions based on the common wisdom of
galaxy formation within the standard cosmological paradigm of $\Lambda$CDM
\citep{boylan23}.
Further observations from NIRSpec and MIRI aboard JWST will allow
testing other predictions of our theory.
We address here the main observable features.

\smallskip  
{\bf Halo masses in {\tt CEERS}.}
We refer in particular to the first {\tt CEERS} sample \citep{finkelstein23},
which consists of 24 galaxies in the range $z \seq 8 \sdash 11.5$ with
apparent magnitudes $29.0 \sdash 26.5$ and 
absolute magnitudes between $M_{\rm 1500} \seq -18.4$ and $-20.6$.
The halo masses for these galaxies can be estimated rather straightforwardly
by ``abundance matching". Specifically, one matches the rank-ordered galaxy 
luminosities to the rank-ordered dark-matter halo masses
in the standard $\Lambda$CDM cosmology. This is done using the theoretical 
halo mass function (HMF) at the given redshift and an estimate of the 
effective survey volume of the sample where it is complete in the given 
luminosity range.

\smallskip
\Fig{hmf} shows the cumulative halo mass function \citep{watson13}
as calibrated by reliable simulations at 
$z \ssim 10$ and above.\footnote{This HMF turns out to be 
rather similar to extrapolations of certain other popular HMFs 
\citep{tinker08}, which were based on simulations at lower redshifts. 
We note that another popular HMF \citep{sheth02}
that is frequently used yield halo number densities that are larger by a 
factor of a few, mostly because the haloes in the calibrating simulation 
were selected based on a friends-of-friends algorithm rather than on 
a spherical volume of overdensity 200.} 
The cumulative HMF provides the comoving number density $n (>\! \Mv)$ 
of haloes with masses above $\Mv$ at different redshifts. 
For a given estimated effective volume $V$ of a given complete sample, 
the expected number of galaxies to be detected above $\Mv$ is 
$N(>\!\!\Mv) = n(>\!\!\Mv)\, V$.
We rank-order the galaxies by their observed absolute luminosities, 
and determine the halo mass of the $i$th galaxy by the inverse of the above 
relation with $N\seq i$.
There is a complication in estimating $V$ because it depends on the selection
magnitude. Using an MCMC method which includes photometric uncertainties,
the estimated effective comoving volume for the brightest {\tt CEERS} galaxies 
is $V \ssim 10^{5} \Mpc^3$ \citep{finkelstein23}.
This yields a halo mass of $\Mv \ssim 10^{11}\msun$ for the brightest galaxy,
which is at $z \ssim 9$.

\smallskip 
{\bf SFE - mass and redshift dependence.}
The five brightest galaxies in the first {\tt CEERS} sample, 
which are at $z \ssim 9$ and thus, based on \fig{hmf}, are expected to 
reside in haloes of $\Mv \sgeq 10^{10.8}\msun$, 
are predicted based on \equ{nsh} to have a 
post-shock density of $\nsh \sgeq 3.5 \stimes 10^3\cmc$.
With enhanced clumpiness, 
this is expected to be sufficient for FFBs,
leading to $\epsilon \ssim 1$.
With $\Ms \seq \eps \fb \Mv$, the predicted stellar masses for the five 
brightest galaxies are $\Ms \sgeq 10^{10}\msun$.
Similarly, expressing SFR $\seq \eps \Mdotac$ with $\eps \ssim 1$,
we predict using \equ{Mdotac} SFR $\sgeq 65 \msun \yr^{-1}$ 
for these brightest galaxies.
Lower mass galaxies at $z \ssim 10$, or galaxies at lower redshifts, 
are predicted by \equ{nsh} and \fig{den} to have densities below $10^3\cmc$, 
corresponding to free-fall times longer than the $1\Myr$ required for
FFBs, and thus to have inefficient star formation with 
$\eps \sll 1$.
The predicted thresholds in $z$ and $\Mv$ for FFB,
as summarized in \fig{mz}, indicate low SFE at $z \slt 8$,
and high SFE at higher redshifts. The high SFE is 
limited to above a mass threshold that decreases steeply with redshift, from 
$\Mv\sgt 10^{10.8}\msun$ at $z \ssimeq 9$
to $\Mv \ssim 3.6\times 10^9\msun$ at $z \seq 15$
and to $\Mv \ssim 6\times 10^8\msun$ at $z \seq 20$.
These predictions are to be compared to stellar masses based on SED fitting
for assumed IMF and star-formation history, which are expected to become
available soon from {\tt CEERS} and other surveys of JWST.

\smallskip 
{\bf Effective stellar radii.}
The shell radius estimate of $\Rsh \slsim 1 \kpc$ in \equ{Rsh_ram},
and the similar estimates for the disc radius,
imply that these galaxies should be compact.
Indeed, the galaxies in the first {\tt CEERS} sample are rather compact, 
with stellar effective radii in the range $\Re \ssim 0.2\sdash 1\kpc$, 
and a median near $\Re \ssim 0.3\sdash 0.4\kpc$ 
\citep[][Fig.~8]{finkelstein23}.

\smallskip 
{\bf Stellar density.}
Using these effective radii, we see in \equ{nobs}
that for the halo masses of the five brightest galaxies 
the average density over the galaxy is $\nobs \ssim \eps\,2\stimes 10^3\cmc$.
If $\eps \ssim 1$, this is in the range that allows FFBs.
At lower halo masses by a factor of a few, $\nobs$ becomes smaller than
$10^3\cmc$ in \equ{nobs}, which is not expected to allow $\eps \ssim 1$.
This is consistent with our predictions based on $\nsh$.

\smallskip 
{\bf Period with high SFE.}
A galaxy that forms its stars in an efficient burst of $\eps \ssim 1$
will be observable with such a high efficiency as long as its halo has not
grown significantly after the FFB phase.
In the EdS regime ($z \sgt 1$),
the typical cosmological halo doubling time is \citep{dekel13}
\be 
M/\dot{M} \ssim 170\Myr \, \Mveight^{-0.14}\, (1+z)_{10}^{-5/2} \, .
\ee
With $\Mv \ssim 10^{10.8}\msun$,
this is roughly the time interval between $z\seq 10$ ($t\ssim 460\Myr$)
and $z\seq 8$ ($t \ssim 630\Myr$).

\smallskip 
{\bf Morphology.}
The robust idea of $\eps\ssim 1$ by FFBs does not rely on the global  
morphology.
According to one of our limiting scenarios, the morphology of the massive
galaxies could be of a thick compact disc with giant clumps. 
In the other scenario addressed, the galaxies could be compact spheroidals 
possibly showing a trace of shell structure.
In fact, a shock at $\Rsh$ may form also in the disc scenario, making a
combined morphological scenario possible.

\smallskip %
{\bf Metallicity, IMF.}
The metallicities are predicted to be relatively low for the stellar winds
not to suppress star formation, $Z \ssim 0.1\Zsun$ and below, 
but not necessarily much lower. 
This may allow a somewhat top-heavy IMF, but a very top-heavy
IMF is not required for $\eps \ssim 1$.

\smallskip
{\bf Gas density.}
The FFB scenario predicts gas densities of $\sim\!2\stimes 10^3\cmc$ and above
in the early stages of the starbursts.
A recent estimate of the electron density in a $z \seq 8.7$ AGN in
{\tt JWST/CEERS} is indeed in the predicted ball park \citep{larson23}.
On the other hand,
the gas fraction within the star clusters at the end of the FFB phase is 
expected to be low, at the level of $\sim\!10\%$ and in a thin disc, 
due to the efficient star formation and the little ejection by radiative 
feedback 
\citep{menon23}.
The post-FFB supernova and stellar-wind ejecta will freely escape from the 
clusters as most of the solid angle is with near-zero column density.
In order to evaluate to what extent the residual inter-cluster gas
will be removed from the galaxy, we 
compare the feedback momentum injection rate of
$\la \dot{p}/m_\star \ra M_\star$ to the gravitational force
$G\, M_{\rm tot}\, M_{\rm gas}\,/\,R^2$.
Assuming that the inter-cluster gas is a fraction $\fg \ssim 0.3$ of the total 
baryonic mass, we learn that ejection is possible once
\be
G\,M_{\rm tot}\,\fgs / R < \la \dot{p}/m_\star \ra \, ,
\label{eq:ejection}
\ee
where $\fgs \seq \Mg/\Ms \seq (\fg^{-1}-1)^{-1}$.
For supernovae, the right hand side can be estimated as
$\la \dot{p}/m_\star \ra \ssim 10^3\kms/40\Myr 
\ssim 8 \stimes 10^{-8} \cm\,{\rm s}^{-2}$.
For $\Ms \ssim 10^{10}\msun$ within $R \ssim 1\kpc$,
the l.h.s. is $\sim\!6\stimes 10^{-8} \cm\,{\rm s}^{-2}$, 
comparable to the r.h.s.,
namely marginally allowing gas ejection from the galaxy.

\smallskip 
{\bf Dust attenuation and color.} 
The dust attenuation in the $z\ssim 10$ FFB galaxies is predicted to be
low, which would make the galaxies blue and as bright as they could be
\citep{ferrara22,ferrara23,ziparo23,fiore23}
once the SFE is high.
Since most of the dust is produced by supernovae, one expects no significant 
dust production during the feedback-free phase itself, in the clusters 
when and where the SFR is high. Therefore, most of the light, which is 
expected to be emitted from the active FFB clusters, is predicted to
be blue with only little dust attenuation.
In addition, 
the dust that is produced by supernovae in post-FFB clusters is expected to be
efficiently removed by the supernova ejecta themselves, as most of the solid 
angle is of near-zero column density 
\citep{menon23}.  
Even if there was significant residual gas spread in the cluster, 
efficient dust ejection was expected based on the momentum balance in 
\equ{ejection}.

\smallskip
{\bf Outflows, inflows, CGM.}
Outflows are expected to be weaker than at lower redshifts
and possibly limited to relatively small radii, with only little extended 
hot CGM.
In particular, outflows with mass-loading factors at the level
of $\sim\!10 \sdash 20\%$ of the common values at lower redshifts 
are expected to be driven by feedback out of the FFB clusters
after most of the gas has been
consumed into stars 
\citep{grudic18,menon23}. 
Intense, cold inflowing streams of radius $\lsim\!1\kpc$ are predicted on
galactic scales.


\section{Discussion}
\label{sec:discussion}

We briefly discuss here certain possible caveats and topics for future 
exploration.

\subsection{Other feedbacks, metallicity, magnetic fields}
\label{sec:other_feedbacks}

Beyond the stellar feedback mechanisms discussed in \se{fbk},
by supernovae, stellar winds, radiative pressure and photoionization,
one should consider whether other sources of feedback or other physical
processes could potentially invalidate the feedback-free period, e.g., 
by stretching the star formation over several free-fall times. 
As summarized above and below, 
the current indications are that all such feedbacks and processes
are ineffective at the low metallicities and high densities and masses proposed
for the FFB star clusters. 

\smallskip 
Proto-stellar jets are such a potential source of suppression of 
SFE within the first few free-fall times, 
but the efficiency increases significantly within $\sim\!3$ free-fall times
\citep{appel22,appel23}.
Furthermore, at cluster masses above $10^4\msun$, the shocks produced by the
protostellar outflows do not have enough energy to escape the clump
(with an escape velocity of $\sim\!20\kms$), so protostellar outflows become
ineffective in suppressing star formation \citep{matzner00}. 
If the efficient star formation is indeed delayed by
$\sim\!3\tff$, the condition for feedback-free starburst will be somewhat more
demanding than assumed above, namely $3\tff \slt \tfbk$. 
However, since we conservatively adopted $\tfbk \ssim 1\Myr$ while the delay
in the feedback is a factor of $2\sdash 3$ times larger, the net effect
on the critical density in \equ{nfbk} would be small.

\smallskip 
The UV background, external or internal to the forming galaxy, is probably not
a significant feedback mechanism at $z \ssim 10$, prior to full re-ionization,
when the star formation is only starting. 

\smallskip 
The possible role of AGN feedback during the proposed FFB phase in
each cluster is yet to be studied. On one hand, the accretion onto the black 
hole may be suppressed by the efficient consumption of gas into stars.
On the other hand, the absence of stellar feedback may allow more efficient
accretion of the residual gas onto the BH.
Intermediate-mass BH seeds of $\sim\!10^3\msun$ may form due to core-collapse 
within the FFB phase, boosted by dynamical friction on the massive stars 
that live for $\sim\!3\Myr$ \citep{portegies02,devecchi09}. These seed BHs can 
grow further on a longer timescale, by gas and stellar accretion and by  
mergers among the ten thousand clusters within the compact galaxy, 
which may eventually lead to massive black holes and AGN feedback 
in the post-FFB phase.

\smallskip 
Another concern is that supernovae from earlier generations of star formation 
may enrich the gas in the star-forming clouds of a later generation, possibly
suppressing the FFB via stellar winds if the metallicity rises well above 
$Z \ssim 0.1\Zsun$.   
Based on the estimate that the cloud destruction time $\tcc$ is much longer 
than the FFB timescales $\tfbk$, $\tff$ and $\tcool$,
we learn that any metals produced in the previous generations and carried
by supernova winds are not going to significantly mix with the shielded 
star-forming gas.
Furthermore, the star-forming gas is expected to be diluted by the incoming 
fresh streams, which are not contaminated by metals as the supernova ejecta 
are limited to the winds that do not penetrate the FFB clouds.

\smallskip 
Furthermore, 
one wishes to evaluate whether magnetic fields may affect the validity 
of the FFB process.
Assuming that during the cloud collapse the magnetic field is  
frozen and it contracts with the gas, and that there is no magnetic enhancement
due to turbulence during the collapse, one can estimate that the magnetic 
energy density, $E_B \seq B^2/(8\pi)$,
will grow from its pre-collapse value of $B_0^2/(8\pi)$ by
a factor $(\rho/\rho_0)^{4/3} \seq \mathcal{M}^{8/3}$,
where $\rho$ and $\rho_0$ are the post-collapse and pre-collapse densities
and $\mathcal{M}$ is the Mach number associated with the collapse. 
While the cosmological evolution of magnetic fields and their value at 
$z \ssim 10$ are still largely unconstrained \citep{subramanian16},
one may assume a typical value of $B_0 \ssim 10^{-9}G$ \citep{mtchedlidze22}.
Comparing the corresponding magnetic energy density to the thermal energy 
density, $E_{\rm therm} \seq n k_{\rm B} T$, the ratio of
magnetic to thermal energy density turns out to be  only $\sim 0.01$ 
even for a Mach number as high as $\mathcal{M} \seq 20$. 
We conclude that the magnetic pressure is not likely to have a significant
effect on the formation of clumps.
We note, however, that the magnetic fields may introduce non-trivial 
corrections to the Jeans stability analysis and to the cooling at low
temperatures, to be studied using MHD simulations with
varying initial conditions for the magnetic fields within the cosmic web.

\subsection{Turbulence}
\label{sec:turbulence}

Another possible concern is that turbulence in the star-forming clouds could 
delay their collapse beyond $\tfbk$ and thus invalidate the FFB conditions. 
We do not consider this likely, for the following reasons.

\smallskip 
Any initial turbulence present in the cloud prior to the start of collapse is 
expected to dissipate in a cloud crossing time \citep{stone98} and therefore 
not to delay the collapse by more than a free fall time. 
This has been confirmed in cloud collapse simulations over a wide range of
initial cloud properties ranging from low-$z$ giant molecular clouds (GMCs)
to high-$z$ metal-poor clouds \citep{fernandez18,grudic21}. 
This would imply that the cloud collapse time is comparable to 
$\tff$, which is $\lsim\! \tfbk$, so the burst should still be feedback-free. 
The gravitational collapse itself could in principle drive additional 
turbulence \citep{federrath11}, but since this is sourced from
the gravitational potential energy itself it cannot prevent global collapse.

\smallskip 
The only way for turbulence to significantly delay the cloud collapse 
is if it is continuously driven. One potential way to drive turbulence in GMCs 
and in the inter-stellar medium (ISM) in general is stellar feedback
\citep{faucher13,hayward17,krum18}, 
which is ineffective by definition in the FFB phase.
A second potential driver of ISM turbulence is torques exerted
by the perturbations associated with violent disc instability, which is
energized by the accompanying inward radial transport 
\citep{dsc09,krum18,ginzburg22}. 
This is unlikely to significantly delay the cloud collapse because
disc instability operates on a disc crossing time, which is much longer than 
the crossing time of the clumps once they become over-dense and start 
collapsing. 
A third potential driver of turbulence is the accreting cold streams.
For massive galaxies at high-$z$ it has been shown that this can be
the primary driver of turbulence in the disc ISM \citep{ginzburg22}. 
However, the efficiency of accretion-driven turbulence is
inversely proportional to the density contrast between the ISM and the
accreting gas \citep{klessen10}. 
While streams may still allow efficient driving of turbulence in the typical 
ISM of high-$z$ discs, where the stream density in the inner halo and in stream
clumps can be comparable to the disc ISM density 
\citep{mandelker18,mandelker20,ginzburg22}, 
once the disc clumps begin collapsing their overdensity grows to the level
where the accretion is unlikely to drive strong turbulence in them. 

\smallskip
A final concern is that the initial turbulence in the pre-collapse cloud might 
have been strong enough to generate large internal density fluctuations which 
collapse faster than the cloud as a whole. This could potentially 
result in the FFBs consisting of many sub-clouds rather than a single 
monolithic cloud. However, the sub-clouds are likely to merge because the 
collapse time of the entire cloud is shorter than $\tfbk$. Sub-cloud 
coagulation is seen in simulations where the velocity field around cooling 
clouds creates an inward pull that causes coagulation \citep{gronke_oh22}.
Thus, the cloud should still collapse without any significant gas dispersal 
or other feedback effects from the early-collapsing sub-clumps. 
Furthermore, such supersonic turbulence is unlikely in the initial clouds 
of $T\ssim 10^4$K, both due to the higher temperature and sound speed and the
more efficient cooling compared to low-$z$ GMCs with $T\ssim 100$K.
This relates to our previous assumption that the bulk kinetic energy
of the streams that hit the central disc/shell, which is at 
Mach$\sim\! 15$, is converted into heat which is quickly radiated away in an
isothermal shock at $T\sim 10^4$K. This leaves fairly little energy
leftover to drive highly supersonic turbulence.

\subsection{Simulations}

The FFB scenario is yet to be verified and quantified via adequate simulations
with accurate subgrid models that can resolve the various FFB phenomena.
Meanwhile, one can gather hints from existing cosmological simulations.
For example, the SIMBA simulations \citep{dave19},
which like other simulations fail to match the abundance of bright galaxies 
at $z\ssim 10$, 
seem to produce an order of magnitude more galaxies in better agreement 
with the indicated observed excess when all the feedback mechanisms are 
artificially turned off (Romeel Dave, private communication).
In another example, cosmological simulations based on the FIRE-2 code
\citep{hopkins18},
where the star formation was artificially limited to above a threshold density 
of $\sim\!10^3\cmc$, similar to the FFB density,
seem to produce massive compact galaxies with high star-formation efficiencies
and certain other similarities to the predicted FFB galaxies 
\citep[][and Robert Feldmann, private communication]{bassini22}.
One should also mention that an empirical model by \citet{behroozi15}, 
in which practically no feedback is assumed, seems to recover the high 
luminosity function at high redshifts \citep[][Fig.~14]{finkelstein23}.

\subsection{Post-FFB evolution}

Certain implications of the FFB scenario are yet to be explored.
Among them is the exciting possibility that the FFB clusters may be potential 
sites for the growth of intermediate-mass black-hole seeds during and 
immediately following the FFB phase.
The presence of massive stars for $\sim\!3\Myr$ in the dense centers of the
young clusters may allow core collapse on a similar timescale, boosted by
the dynamical friction that brings the massive stars in, which may lead to 
central black holes of mass larger than $\sim 10^{-3}$ of the cluster mass  
\citep{portegies02,devecchi09,devecchi12}, namely $\Mbh\ssim 10^3\msun$.
These black holes will grow further on a longer timescale by collisions with 
stars \citep{stone17,rizzuto23} and gas accretion \citep{schleicher22}.
Then, the mergers and interactions between the thousands clusters in the 
compact $<\!\kpc$ galaxies may stimulate further accretion and black-hole 
mergers into more massive ones, en route to 
super-massive black holes at high redshifts. 

\smallskip 
Another implication is the formation and survival of globular clusters.
The typical stellar density in today's globular clusters is $\sim 10^4\cmc$
(e.g. assuming a stellar mass of $2.5\times 10^5\msun$ 
with a half-mass radius $5\pc$. 
This is indeed in the ball park of the FFB densities,
possibly indicating that the oldest globular clusters were formed in
the $z\sim 10$ FFB galaxies.
The $z \seq 0$ descendents of the haloes hosting the FFB brightest galaxies
at $z\ssim 10$ are expected to be galaxy-cluster haloes of
$2\stimes 10^{14}\msun$.
This is assuming a halo mass growth by $e^{0.8 z}$ from $z$ until
the present \citep{dekel13}.
The descendants of the brightest $z \ssim 10$ galaxies are thus 
central galaxies in galaxy clusters.
Remnants of the original star clusters that used to
constitute the whole FFB galaxy may remain as a population of globular 
clusters in today's brightest cluster galaxies.
In the extreme case where all the FFB clusters survived, and only them, 
today's ratio of mass in globular clusters and in haloes would be 
$10^{10}/10^{14.3} \ssim 10^{-4.3}$, comparable to the standard observational
value \citep{boylan17}. This indicates that the FFB scenario does not predict 
an excessive population of stars in very dense environments today.
The evolution of this cluster population, as well as the possible relevance of
FFBs to other types of clusters, are yet to be explored.

\subsection{The role of mergers}

The possible role of mergers in $z \ssim 10$ FFBs will be worth considering.
A major merger at $z \seq 0$ typically generates a SFR peak of duration 
$\Delta t \ssim 100\Myr$ or longer \citep{hopkins13}.  
This is less than a percent of the Hubble time, 
which would translate to $3.5\Myr (1+z)_{10}^{-3/2}$ at $z$.
At $z \ssim 10$ this is somewhat longer than $\tfbk$, making the mergers less 
likely as the sources of feedback-free star formation at that epoch,
but their role is not ruled out at higher redshifts.
The time between major mergers of haloes is expected to be in the ball park of
the mass doubling time by accretion, which is very crudely
\citep{neistein08,dekel13}
\be
\tmer \ssim 100\Myr\, (1+z)_{10}^{-5/2} \, .
\ee
This implies that a typical FFB galaxy had on the order of one such merger 
during the duration of its formation at $z \ssim 10$.
The scenario of a FFB due to a merger would therefore involve a monolithic 
collapse with a single global starburst.


\smallskip
For $z \slt 7$, where $\eps \ssim 0.1$, and in the galactic mass range
where $\Ms/\Mv$ is a growing function of $\Mv$,
the galaxies in haloes of masses below $\Mv \ssim 10^{11}\msun$
are not expected to have significant long-lived gas disc components
\citep{dekel20_flip}.
This has been seen in simulations and was demonstrated to emerge
from the fact that below this critical mass the typical time interval between
destructive major mergers, as defined by the baryonic mass ratio rather than
the halo mass ratio, is shorter than the gas disc orbital time.
However, with $\eps \ssim 1$, the baryonic merger timescale becomes largely
mass independent, like the halo merger timescale.
Since the orbital time is also mass independent, 
and since the merger timescale is longer than the orbital time at $z \ssim 10$,
gas discs may be possible over a broader mass range.
The role of mergers in FFBs and disc survival are yet to be explored.

\section{Conclusion}
\label{sec:conc}

We predict high-efficiency conversion of accreted gas to stars due to 
feedback-free starbursts (FFBs) in the brightest galaxies at $z\ssim 10$ and
above, 
with $\Ms \ssim \fb \Mv$ and SFR comparable to the baryon accretion rate.
The key is a characteristic density of a few $\times\!10^3\cmc$ in 
star-forming clouds, corresponding to a free-fall time of $\ssim 1\Myr$. 
At the relatively low metallicities expected at these redshifts,
this free-fall time is shorter than the delay between a starburst and 
the onset of effective stellar winds and supernovae feedback.
At the corresponding high surface density in the star-forming clouds,
the radiative feedback is also incapable of significantly suppressing the 
star formation.
At higher densities, the timescale for cooling below $10^4$K becomes shorter 
than the free-fall time, allowing starbursts at $\sim 10$K. 

\smallskip 
The key elements of FFB galaxy formation at the high densities and low
metallicities expected at $z \ssim 10$ are as follows.

\no\bul
There is a delay of $\tfbk \sgsim 1 \Myr$ in the onset of supernova feedback
after a starburst, which is also valid for solar winds when the
metallicity is $Z \sleq 0.2\Zsolar$.

\no\bul
For densities above a threshold $\nfbk \ssim 2\stimes 10^3\cmc$, 
the free-fall time
is shorter than $1\Myr$, allowing feedback-free star formation.

\no\bul
The surface density in clouds of density $n\sgeq 2\stimes 10^3\cmc$ and
radius $\sim\!15\pc$ is $\Sigma \sgeq 2\stimes 10^3\Mpc\pc^{-2}$, where 
radiative feedback is also ineffective in suppressing star formation. 

\no\bul
For comparable and somewhat higher densities, the cooling time at 
$T \sleq 10^4$K down to star-formation temperatures is also comparable 
to the free-fall time, thus triggering short starbursts.

\no\bul
Clouds of masses $>\! 10^4\msun$ are shielded against winds
and radiation from earlier generations of stars.

\no\bul
At $z \ssim 10$ and $\Mv \ssim 10^{10.8}\msun$ and beyond, 
the characteristic density of a few $\times 10^3\cmc$ that 
permits FFBs {\it emerges naturally}.
It can be derived from the inner post-shock density, or in 
shells or rotating discs that form at these epochs.

\no\bul
Low SFE values are expected at $z \slt 8$.
High SFE values due to FFB are expected at higher redshifts,
above a mass threshold
$\Mv \ssim 10^{10.5}\msun$ at $z \sim 10$, which drops to  
$\Mv \ssim 3.5\times 10^9\msun$ at $z \ssim 15$. 

\no\bul
The FFB starbursts are expected in dense clouds that fragment 
from a shell or a disc, in which the density surpasses the threshold density 
for FFB and the surface density is sufficiently
high for radiative feedback to be ineffective. 
Most of the star formation is expected to be in 
clouds that are sufficiently massive for shielding against feedback from 
other clusters.

\no\bul
A necessary condition for FFBs is that
the haloes at $z \ssim 10$ are in the cold-flow regime, where the atomic cooling
time is much shorter than the halo free-fall time. The gas accreted onto the
halo flows in efficiently via $\sim\!10^4$K streams of radii $\lsim\! 1\kpc$, 
which gradually feed the central galaxy over $\tvir \ssim 80\Myr$, allowing
several generations of starbursts in star clusters.

\no\bul
The galaxies are expected to be compact, with radii of $\Rsh \slsim 1 \kpc$.

\smallskip 
The observable predictions can be summarized as follows.

\no\bul
For the most massive galaxies of halo mass $\Mv \ssim 10^{10.8}\msun$ 
at $z \ssim 10$, we predict a typical stellar density of 
$n \ssim 3 \stimes 10^3\cmc$, 
with an SFE $\eps \ssim 1$.
This implies stellar masses of $\Ms \ssim 10^{10}\msun$
and star-formation rates of $\sim\! 65\Msun\yr^{-1}$
inside galactic effective radii of $\Re \slsim 1\kpc$.
At lower redshifts or in haloes of lower masses the efficiency $\eps$
is expected to be lower.
At higher redshifts the mass threshold is expected to be lower.

\no\bul
The morphology of the massive galaxies could be of a thick compact disc 
with giant clumps, while the lower-mass galaxies could be smoother discs or 
non-discs.
Alternatively, the star formation is predicted to occur in compact spheroidal
shells.

\no\bul
The metallicities are required to be relatively low, below $Z \ssim 0.2$,
but not necessarily much lower.

\no\bul
Because of the feedback-free starbursts,
the dust attenuation is expected to be weak,
and the post-FFB gas fractions are expected to be relatively low. 

\no\bul
Any outflows are expected to be relatively weak and likely
limited to relatively small radii, with only little extended hot CGM.

\smallskip
The implications for the post-FFB phase are yet to be studied.
One implication is possible survival of a fraction of the FFB clusters
as globular clusters in today's brightest cluster galaxies.
Another possible implication is efficient formation of massive black holes,
starting with seed BHs by core collapse within the clusters, boosted by 
the dynamical friction of the massive stars of lifetimes $\sim\!3\Myr$
within the FFB clusters, 
and continuing with gas and stellar accretion and the cluster-cluster 
interactions in the compact galaxies.
A third implication is the potential effect on the cosmological
reionization process. 

\smallskip
The FFB phenomenon is yet to be verified and quantified through proper 
simulations with accurate subgrid models that can capture the processes 
involved in the FFBs.


\section*{Acknowledgments}
We are grateful for stimulating discussions with Michael Boylan-Kolchin,
Romeel Dave, Dhruba Dutta-Chowdhury,
Andrea Ferrara, Steve Finkelstein, Claude-Andre Faucher-Giguere, 
Omri Ginzburg, Mike Grudic, Martin Haehnelt, 
Andrey Kravtsov, Mark Krumholz, Chris McKee, 
Shyam Menon, Volker Springel, Nicolas Stone, Jeremy Tinker, Marta Volonteri 
and Zhiyuan Yao.
This work was supported by the Israel Science Foundation Grants
ISF 861/20 (AD, ZL), 2190/20 (KCS), 2190/20 (YB) and 3061/21 (NM, ZL),
and by the DFG/DIP grant STE1869/2-1 GE625/17-1 (AD, KCS, ZL).

\section*{DATA AVAILABILITY}

Data and results underlying
this article will be shared on reasonable request to the corresponding author.



\bibliographystyle{mnras} 
\bibliography{z10}

\label{lastpage}
\end{document}